\documentclass[8.5pt,twoside,twocolumn]{article}
\usepackage[super,sort&compress,comma]{natbib} 
\usepackage{mhchem}
\usepackage{times,mathptmx}
\usepackage{sectsty}
\usepackage{balance} 

\usepackage{graphicx} 
\usepackage{lastpage}
\usepackage[format=plain,justification=raggedright,singlelinecheck=false,font=small,labelfont=bf,labelsep=space]{caption} 
\usepackage{fancyhdr}
\usepackage{color}
\usepackage{amssymb}
\usepackage{amsbsy}
\begin{document}

\thispagestyle{plain}
\fancypagestyle{plain}{
\fancyhead[L]{\includegraphics[height=8pt]{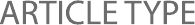}}
\fancyhead[C]{\hspace{-1cm}\includegraphics[height=20pt]{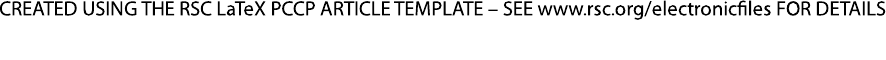}}
\fancyhead[R]{\includegraphics[height=10pt]{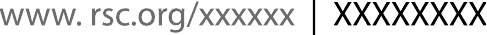}\vspace{-0.2cm}}
\renewcommand{\headrulewidth}{1pt}}
\renewcommand{\thefootnote}{\fnsymbol{footnote}}
\renewcommand\footnoterule{\vspace*{1pt}%
\hrule width 3.4in height 0.4pt \vspace*{5pt}} 
\setcounter{secnumdepth}{5}

\makeatletter 
\def\subsubsection{\@startsection{subsubsection}{3}{10pt}{-1.25ex plus -1ex minus -.1ex}{0ex plus 0ex}{\normalsize\bf}} 
\def\paragraph{\@startsection{paragraph}{4}{10pt}{-1.25ex plus -1ex minus -.1ex}{0ex plus 0ex}{\normalsize\textit}} 
\renewcommand\@biblabel[1]{#1}            
\renewcommand\@makefntext[1]%
{\noindent\makebox[0pt][r]{\@thefnmark\,}#1}
\makeatother 
\renewcommand{\figurename}{\small{Fig.}~}
\sectionfont{\large}
\subsectionfont{\normalsize} 

\fancyfoot{}
\fancyfoot[LO,RE]{\vspace{-7pt}\includegraphics[height=9pt]{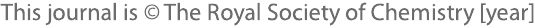}}
\fancyfoot[CO]{\vspace{-7.2pt}\hspace{12.2cm}\includegraphics{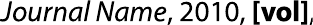}}
\fancyfoot[CE]{\vspace{-7.5pt}\hspace{-13.5cm}\includegraphics{RF}}
\fancyfoot[RO]{\footnotesize{\sffamily{1--\pageref{LastPage} ~\textbar  \hspace{2pt}\thepage}}}
\fancyfoot[LE]{\footnotesize{\sffamily{\thepage~\textbar\hspace{3.45cm} 1--\pageref{LastPage}}}}
\fancyhead{}
\renewcommand{\headrulewidth}{1pt} 
\renewcommand{\footrulewidth}{1pt}
\setlength{\arrayrulewidth}{1pt}
\setlength{\columnsep}{6.5mm}
\setlength\bibsep{1pt}

\twocolumn[
  \begin{@twocolumnfalse}
\noindent\LARGE{\textbf{Tunable structures of mixtures of magnetic particles in liquid-crystalline matrices}}
\vspace{0.6cm}

\noindent\large{\textbf{Stavros D. Peroukidis,$^{\ast}$\textit{$^{a}$} Ken Lichtner,\textit{$^{a}$} and
Sabine H.L. Klapp\textit{$^{a}$}}}\vspace{0.5cm}

\noindent\textit{\small{\textbf{Received Xth XXXXXXXXXX 20XX, Accepted Xth XXXXXXXXX 20XX\newline
First published on the web Xth XXXXXXXXXX 200X}}}

\noindent \textbf{\small{DOI: 10.1039/b000000x}}
\vspace{0.6cm}

\noindent \normalsize{We investigate the self-organization of a binary mixture of similar sized rods and dipolar soft spheres by means of Monte-Carlo simulations. 
We model the interparticle interactions by employing anisotropic Gay-Berne, dipolar and soft-sphere interactions. In the limit of vanishing magnetic moments we obtain a variety of fully miscible liquid crystalline phases including nematic, smectic and lamellar phases. 
For the magnetic mixture, 
we find that the liquid crystalline matrix supports the formation of orientationally ordered ferromagnetic chains.   
Depending on the relative size of the species the chains align parallel or perpendicular to the director of the rods forming uniaxial or biaxial nematic, smectic and lamellar phases. As an exemplary external perturbation we apply a homogeneous magnetic field causing uniaxial or biaxial ordering to an otherwise isotropic state.  }
\vspace{0.5cm}
 \end{@twocolumnfalse}
  ]



\footnotetext{\textit{$^{a}$~Institute of Theoretical Physics, Technical University Berlin, Secr. EW 7-1 Hardenbergstr. 36, D-10623 Berlin, Germany. Fax: +49 30 314-21130; Tel:+49 30 314-28851; E-mail: peroukid@mailbox.tu-berlin.de}}



\section{Introduction}
In the last decades significant progress has been
made in understanding the phase behavior of mixtures of rods and spheres. 
The great interest in these systems was coined by the pioneering work of Asakura and Oosawa in 1954 \cite{Asakura54} in which 
they predicted that an effective attractive force arises between two large particles in a suspension of considerable smaller depletion agents. As a consequence, even the simplest mixtures composed of purely repulsive rods and spheres exhibit a rich phase behavior \cite{Lekkerkerker99}. In the subsequent period, phase and reentrant transitions as well as multiphase equilibria between gas, liquid, liquid crystalline (LC) and crystal phases have been observed by investigations based on experiments \cite{Fraden98,Jeroen10}, theoretical studies \cite{Lekkerkerker99,Schmidt02,PeroukidisMix,Cinacchi04} and computer simulations \cite{Bolhuis03,Antypov04,Urakami03,Dogic2000}. 

Correspondingly, even more complex behavior is expected for suspended particles with internal degrees of freedom such as magnetic particles. As predicted in a  famous work by Brochard and de Gennes the coupling between the ferrocolloids and the liquid crystal molecules increases the sensitivity to  external magnetic field significantly \cite{degennes70,Chen83}; a phenomenon with many potential applications for magneto- \cite{Mirantsev14,Mertelj14} or electro-optical devices \cite{Kreuzer92,Kreuzer93}. So far, suspensions of ferrocolloids in a nematic LC matrix composed of much smaller LC molecules (also referred to as \textit{low molecular mass} LCs) have been studied experimentally and theoretically (see, e.g., Refs.~[\negthickspace\citenum{Kopacansky2008,Buluy2011}] and references therein).

At the present time, increasing attention is being paid to the study of mixtures of spherical magnetic and rodlike colloids where the spheres are of similar size as compared to the width of the rods. 
Recent experimental examples include magnetic nanoparticle suspensions \cite{Kredentser13}, attempts \cite{privcomm} to produce suspensions of magnetic nanoparticles in dichroic pigment particles \cite{May14} and percolating carbon nanotubes \cite{Schoot11}.
 
In the present study, we consider such a colloidal rod-sphere mixture with additional permanent 
dipole moments for the spheres. 
Dipoles exhibit a long-ranged  dipole-dipole interaction 
with the head-to-tail alignment being the most energetically favorable spatial configuration. In fact, dipolar particles alone show 
even in the absence of an external (electric or) magnetic field complex microstructure formation including 
chain-like \cite{Klapp05}, network-like \cite{Ilg13}, and ring-like structures \cite{Sciortino11,Sciortino12}. 
Prime examples of self-assembled dipolar structures fueled by external fields include chain formation in constant fields \cite{Martin98}, layer formation in rotating fields \cite{Sebastian11}, and the structure formation in triaxial
fields \cite{Douglas10}
with a wide range of applications in magnetorheology \cite{odenbachbook} and 
biomedicine \cite{Lewis15,Wu12}.

The aim of the present study is three-fold: firstly, to set-up a tractable model for rod-(nondipolar) sphere mixtures and explore the corresponding LC order, secondly, to provide information to which extend both types of ordering, i.e. ferromagnetic and LC ordering, can be combined in mixtures involving dipolar interactions, and thirdly, to illuminate how far the collective behavior can be manipulated by coupling an external magnetic field to the dipolar spheres.

To our knowledge, no theoretical investigations have been carried out on the particle level for such a magnetic hybrid system and thus, a microscopic understanding, is so far missing. As a first step, we have recently presented a Monte-Carlo (MC) simulation  study\cite{StavarXiv} targeting two specific size ratios in the absence of any perturbation. In the present paper, we aim to explore in more detail the behavior of the underlying reference system (involving non-magnetic spheres), as well as other relevant size ratios. Moreover, we investigate the 
impact of the simplest external perturbation, that is, an homogeneous external field.

The remainder of the paper is organized as follows. In
Section 2 we formulate the theoretical model and give a brief outline of the simulation details. Section 3 presents results for the rod-sphere mixture. We begin with the non-magnetic system in Section 3.1 followed by a discussion of the dipolar system in Section 3.2. The influence of an external magnetic field is presented in Section 3.3. We conclude in Section 4. 

\section{Model}
Our model fluid consists of a mixture of dipolar soft spheres with embedded, permanent dipole moments and uniaxial rods. The total interaction energy for the system may be decomposed into contributions stemming from each species  and a contribution accounting for interactions between rods (r) and spheres (s), that is, 
\begin{align}
U^\text{int}=\sum\limits_{i=1}^{N_\mathrm{s}}\sum\limits_{j\neq i}^{N_\mathrm{s}}U_{ij}^\text{s}+\sum\limits_{i=1}^{N_\mathrm{r}}\sum\limits_{j\neq i}^{N_\mathrm{r}}U_{ij}^\text{r}+\sum\limits_{i=1}^{N_\mathrm{s}}\sum\limits_{j=1}^{N_\mathrm{r}}U_{ij}^\text{rs}
\end{align}
where $N_\mathrm{s}$ is the number of spheres and $N_\mathrm{r}$ is the number of rods. 

For the interaction between rods we use a single-site potential model suggested by
Gay and Berne\cite{GayBerne81} which considers an orientation-dependent range parameter between pairs of particles. 

The latter reduces in the case of two identical particles of length $l$ and width $\sigma_0$ to \cite{GayBerne81}

\begin{align}
\sigma(\mathbf{\hat{u}}_i,\mathbf{\hat{u}}_j,\hat{\mathbf{r}}_{ij})=\sigma_0&\left(1-\frac{\chi}{2}\left[\dfrac{(\hat{\mathbf{r}}_{ij}\cdot\mathbf{\hat{u}}_i+\hat{\mathbf{r}}_{ij}\cdot\mathbf{\hat{u}}_j)^2}{1+\chi\mathbf{\hat{u}}_{i}\cdot\mathbf{\hat{u}}_j}\right.\right.\nonumber\\
&+\left.\left.\dfrac{(\hat{\mathbf{r}}_{ij}\cdot\mathbf{\hat{u}}_i-\hat{\mathbf{r}}_{ij}\cdot\mathbf{\hat{u}}_j)^2}{1-\chi\mathbf{\hat{u}}_{i}\cdot\mathbf{\hat{u}}_j}
\right]\right)^{-\frac12}\label{Eq.sigma}
\end{align}
where $\chi=(l^2/\sigma_0^2-1)/(l^2/\sigma_0^2+1)$, $\mathbf{\hat{u}}_i$ is the director along the principal axis of particle $i$ and $\mathbf{r}_{ij}$ is the connecting vector between the centre of masses of particles $i$ and $j$. 
With the range parameter $\sigma$ defined in Eq.~(\ref{Eq.sigma}) we can now introduce a modified Gay-Berne (GB) potential of Lennard-Jones form, that is, \cite{Antypov04}
\begin{align}
U_{ij}^\text{r}(\mathbf{\hat{u}}_i,\mathbf{\hat{u}}_j,\mathbf{r}_{ij})=4\varepsilon&(\mathbf{\hat{u}}_i,\mathbf{\hat{u}}_j,\hat{\mathbf{r}}_{ij})\left[\left(\dfrac{\sigma_0}{\lvert\mathbf{r}_{ij}\rvert-\sigma(\mathbf{\hat{u}}_i,\mathbf{\hat{u}}_j,\hat{\mathbf{r}}_{ij})+\sigma_0}\right)^{12}\right.\nonumber\\
&-\left.\left(\dfrac{\sigma_0}{\lvert\mathbf{r}_{ij}\rvert-\sigma(\mathbf{\hat{u}}_i,\mathbf{\hat{u}}_j,\hat{\mathbf{r}}_{ij})+\sigma_0}\right)^{6}\right]\label{Eq.GBLJ}.
\end{align}
For the strength anisotropy parameter in Eq.~(\ref{Eq.GBLJ}) we use \cite{GayBerne81}
\begin{align}
\varepsilon(\mathbf{\hat{u}}_i,\mathbf{\hat{u}}_j,\hat{\mathbf{r}}_{ij})=\varepsilon_0\left[\varepsilon_1(\mathbf{\hat{u}}_i,\mathbf{\hat{u}}_j)\right]^\nu\left[\varepsilon_2(\mathbf{\hat{u}}_i,\mathbf{\hat{u}}_j,\hat{\mathbf{r}}_{ij})\right]^\mu\label{Eq.strparam}
\end{align}
with $\varepsilon_0$, $\varepsilon_1(\mathbf{\hat{u}}_i,\mathbf{\hat{u}}_j)=\left[1-\chi^2(\mathbf{\hat{u}}_{i}\cdot\mathbf{\hat{u}}_j)^2\right]^{-\frac12}$ being the strength parameters from the original formulation of the overlap model by Berne and Pechukas \cite{BernePechukas72} (where $\mu$, $\nu$ are adjustable exponents). The latter parameter in Eq.~(\ref{Eq.strparam}), $\varepsilon_2(\mathbf{\hat{u}}_i,\mathbf{\hat{u}}_j,\hat{\mathbf{r}}_{ij})$, has been introduced later by Gay and Berne \cite{GayBerne81} to adjust the well depth ratio for the side-by-side ($\varepsilon_s$) to end-to-end ($\varepsilon_e$) configuration by introducing the parameter $\chi'=(\varepsilon_s^{1/\mu}-\varepsilon_e^{1/\mu})/(\varepsilon_s^{1/\mu}+\varepsilon_e^{1/\mu})$, that is,
\begin{align}
\varepsilon_2(\mathbf{\hat{u}}_i,\mathbf{\hat{u}}_j,\hat{\mathbf{r}}_{ij})=&1-\dfrac{\chi'}{2}\left[\frac{(\hat{\mathbf{r}}_{ij}\cdot\mathbf{\hat{u}}_{i}+\hat{\mathbf{r}}_{ij}\cdot\mathbf{\hat{u}}_j)^2}{1+\chi'\mathbf{\hat{u}}_{i}\cdot\mathbf{\hat{u}}_{j}}\right.\nonumber\\
&\left.+\frac{(\hat{\mathbf{r}}_{ij}\cdot\mathbf{\hat{u}}_{i}-\hat{\mathbf{r}}_{ij}\cdot\mathbf{\hat{u}}_j)^2}{1-\chi'\mathbf{\hat{u}}_{i}\cdot\mathbf{\hat{u}}_{j}}\right].\label{Eq.eps2}
\end{align}

For the pair potential between two dipolar soft spheres (DSS) of diameter $\sigma_s$ we use 
\begin{align}
U_{ij}^\text{s}(\mathbf{r}_{ij})=u_{\text{ss}}(\mathbf{r}_{ij})+u_{\text{dd}}(\mathbf{r}_{ij})
\end{align}
where $u_{\text{ss}}$ is the (truncated and shifted, cf. \cite{allen}) soft-sphere (SS) repulsion given by
\begin{align}
u_{\text{ss}}(\mathbf{r}_{ij})=4\varepsilon_0\frac{\sigma_s^{12}}{\lvert\mathbf{r}_{ij}\rvert^{12}}
\end{align}
and $u_{\text{dd}}$ accounts for the dipole-dipole potential stemming from the permanent dipole moments $\mathbf{m}_{i}$, that is,
\begin{align}
u_{\text{dd}}(\mathbf{m}_i,\mathbf{m}_j,\mathbf{r}_{ij})=\frac{\mathbf{m}_{i}\cdot \mathbf{m}_{j}}{\lvert\mathbf{r}_{ij}\rvert^{3}}-3\frac{(\mathbf{m}_{i}\cdot\mathbf{r}_{ij})(\mathbf{m}_{j}\cdot\mathbf{r}_{ij})}{\lvert\mathbf{r}_{ij}\rvert^{5}}.
\end{align}

Finally, for the interaction between pairs of rods and spheres we consider a Gay-Berne potential of the same form of Eq.~(\ref{Eq.GBLJ}). To this end, we have to determine the range parameter depending only on the orientation of rod $j$, $\hat{\mathbf{u}}_j$, and the normalized vector $\hat{\mathbf{r}}_{ij}$ connecting to the centre-of-mass of sphere $i$, that is, \cite{Cleaver96}
\begin{align}
\sigma_{\text{rs}}(\hat{\mathbf{u}}_j,\hat{\mathbf{r}}_{ij})=\sigma_0^{\text{rs}}\left[1-\chi''(\hat{\mathbf{r}}_{ij}\cdot\hat{\mathbf{u}}_j)^2\right]^{-\frac{1}{2}}
\end{align}
where $\sigma_0^{\text{rs}}=\frac12 (\sigma_0+\sigma_\text{s})$ and $\chi''=(l^2-\sigma_0^2)/(l^2+\sigma_\text{s}^2)$. 
Similarly, the strength anisotropy for pairs of rods and spheres becomes
\begin{align}
\varepsilon_{\text{rs}}(\hat{\mathbf{u}}_j,\hat{\mathbf{r}}_{ij})=\varepsilon_0\left[1-\chi'''(\hat{\mathbf{r}}_{ij}\cdot\hat{\mathbf{u}}_j)^2\right]^{\mu}
\end{align}
where the rod-sphere well-depth anisotropy is given by $\chi'''=1-(\varepsilon_e/\varepsilon_s)^{1/\mu}$. For our calculations we choose an established set of parameters for which a large number of simulation studies have been carried out, that is, $l/\sigma_0=3$, $\varepsilon_s/\varepsilon_e=5$, $\mu=2$ and $\nu=1$. In the present study the diameter of the spheres $\sigma_\text{s}$ remains as an adjustable parameter with $\sigma^{*}_{\text{s}}=\sigma_{\text{s}}/\sigma_0$. For the numerical computations of the potentials we use cutoff radii of $r_c^\text{r}=4\sigma_0$ for the rod-rod interactions, $r_c^\text{ss}=2.5\sigma_s$ for the short-ranged soft-sphere repulsions, and $r_c^\text{rs}=2(l+\sigma_s)$ for the rod-sphere interactions. The long-ranged dipolar interactions are treated by the Ewald summation method with conducting boundaries \cite{Ewald07}.

We have examined systems for a variety of particle compositions ($\textit{x}_{\text{a}}=\textit{N}_{\text{a}}/\textit{N} $ where $\text{a}=\text{s},\text{r} $ for spheres and rods, respectively), volume fractions ($\phi_{\text{a}}=N_{\text{a}}u_{\text{a}}/V$ where $u_\text{a}$ is a particles' volume), and total number densities $\rho^*=\textit{N}\sigma^3_0/\textit{V} $.  
We have performed Monte-Carlo (MC) simulations \cite{allen} in the canonical ensemble to examine the morphologies of these systems. Equilibration requires of the order of $2$x$10^6$ cycles and a further $5$x$10^5$-$1$x$10^6$ cycles is being used for the calculation of ensemble averages of quantities of interest. A MC cycle consists of \textit{N} trial attempts (moves, orientations and moves-orientations) for a randomly chosen particle.

The orientational ordering of the phases has been quantified with the aid of order parameters. More specifically, we measure the orientational ordering via the nematic order parameter $S^{({\text{a}})}$ obtained through diagonalizing the ordering tensor \cite{Camp1999}
\begin{align}
\boldsymbol{Q}_{{\text{bc}}}^{{\text{a}}}=\sum_{i=1}^{N_{{\text{a}}}}\left[3\left(K_{i}^{{\text{a}}}\right)_{{\text{b}}} \left(K_{i}^{{\text{a}}}\right)_{{\text{c}}}-\delta_{{\text{bc}}}\right]/2N_{{\text{a}}}
\end{align}
where $\text{b,c}=x,y,z$ (cartesian components)  and $K_{i}^{{\text{a}}}$ is the cartesian  component of the symmetry axis of rods or dipolar spheres (along the direction of $\textbf{m}_i$).  For each species, the eigenvector associated with the largest eigenvalue  $S_+ $ of the ordering tensor is considered with respect to the director $\hat{\mathbf{n}}_{\text{s}}$ or $\hat{\mathbf{n}}_{\text{r}}$. The other two eigenvalues, $S_0 $ and $S_-$, fulfill the inequality $S_{+}>S_{0}\geq S_{-}$. The biaxiality of the phase can be calculated via the order parameter \cite{Cuetos2008} 
\begin{align}
B=\left\langle \frac32\left(\hat{\mathbf{n}}_{\text{r}}\cdot\hat{\mathbf{n}}_{\text{s}} \right)^{2}-\frac12\right\rangle
\end{align}
which yields $B=1$ (parallel directors) and $B=-0.5$ (perpendicular directors) for an uniaxial and a biaxial phase, respectively. The polarity of the magnetic phase is measured via the first rank polar order parameter, that is,
\begin{align}
P_{1}=\left\langle \frac{1}{N_{\text{s}}} \left\vert\sum_{i=1}^{N_{{\text{s}}}}\hat{\mathbf{m}}_{i}\cdot\hat{\mathbf{n}}_{{\text{s}}}\right\vert\right\rangle.\label{eq.polar}
\end{align}

An analysis of translational ordering has been performed by means of correlation functions \cite{Veerman1992,Bebo2000,Mcgrother1996}. We have calculated (i) the longitudinal correlation function 
\begin{align}
g_{\Vert;\mathbf{\hat{n}}_{\text{a}}}^{\left({\text{a}}\right)}\left(r_{\Vert}\right)= \left\langle \dfrac{\sum_{i\neq j}\delta\left(r_{\Vert}- \left|\mathbf{r}_{ij}\cdot\hat{\mathbf{n}}_{\text{a}}\right|  \right)}{\Delta V_2\rho\left(N_{\text{a}}-1\right)}   \right\rangle
\end{align}
of the projection of the intermolecular vector parallel to the macroscopic principal director $\hat{\mathbf{n}}_{\text{a}} $ where $\Delta V_2=\pi\left( r^2 -\left(\mathbf{r}_{ij}\cdot\hat{\mathbf{n}}_{\text{a}}\right)^2 \right)\Delta r_{\Vert} $ is the volume of a cylindrical shell with thickness $\Delta r_{\Vert}=0.05$, (ii) the dipole-dipole correlation function perpendicular to the director $\hat{\mathbf{n}}_{\text{a}}$, that is,
\begin{align}
g_{1;\hat{\mathbf{n}}_{\text{a}}}^{\left({\text{a}}\right)}\left(r_{\bot}\right) = \left\langle\dfrac{\sum_{i\neq j}\delta\left(r_{\bot}- \sqrt{r_{ij}^2-\left(\mathbf{r}_{ij}\cdot\hat{\mathbf{n}}_{\text{a}} \right)^{2}}\right)\cos\theta_{ij}   }{\sum_{i\neq j}\delta\left(r_{\bot}- \sqrt{r_{ij}^2-\left(\mathbf{r}_{ij}\cdot\hat{\mathbf{n}}_{\text{a}} \right)^{2}}\right)   }\right\rangle
\end{align}
where $\cos\theta_{ij}= \hat{\mathbf{z}}_{i}\cdot\hat{\mathbf{z}}_{j} $ and $\hat{\mathbf{z}}_{i}$ is the pricipal axis (given by the dipole vector). Finally, to analyze the structure with respect to the dipole moment of a DSS particle we have calculated the two-dimensional correlation function 
\begin{align}
&g^{({\text{s}})}\left(r_{\Vert},r_{\bot}\right)=\nonumber\\
&\left\langle \dfrac{\sum_{i\neq j}\delta\left(r_{\bot}- \sqrt{r_{ij}^2-\left(\mathbf{r}_{ij}\cdot\hat{\mathbf{m}}_{i} \right)^{2}}\right)\delta\left(r_{\Vert}- \left|\mathbf{r}_{ij}\cdot\hat{\mathbf{m}}_{i}\right|  \right) }{\Delta V\rho\left(N_{\text{s}}-1\right)}\right\rangle
\end{align}
where $\Delta V=\pi\left( \left(r_{\bot}+\Delta r_{\bot} \right)^2 - r_{\bot}^2 \right)\Delta r_{\Vert} $  and $\mathbf{\hat{m}}_{i}$ is the dipolar unit vector of particle $i$.
 
\section{Results and Discussion}

\subsection{Binary mixtures of rods and soft spheres (GB-SS)}

In this section we study the self-organization of a binary mixture of Gay-Berne rods and repulsive soft spheres (GB-SS mixture) for different sphere diameters ranging from $\sigma^{*}_{\text{s}}=1.0$  to $\sigma^{*}_{\text{s}}=2.0$. The GB-SS mixtures are considered as reference systems since the particles' self-organization  is essential for the examination of more complex binary mixtures of rods and dipolar soft spheres (GB-DSS). In particular, the configurations of the GB-SS system are considered as initial configurations for the GB-DSS system. We have focused on systems with $\textit{x}_{\text{r}}=0.8$ and $\textit{x}_{\text{r}}=0.9$ for which fully miscible phases are obtained. This is not a trivial finding since the amounts of spheres that can be supported by LC phases depends crucially on the chemical affinity of the species \cite{PeroukidisMix}. Hence, binary mixtures of (GB) rods and spherical particles which attract each other, such as Lennard-Jones spheres, exhibit demixing transitions \cite{Antypov04} or microphase separation occurs in hard-core interacting rod-sphere mixtures \citep{Dogic2000}.  Depending on the size of the spheres these values correspond to volume fraction ratios (rods to spheres)  $2.4<\phi_{\text{r}}/\phi_{\text{s}}<12$ which is of the order of magnitude that can be achieved in real colloidal suspensions \cite{privcomm}.

\begin{center}
  \begin{figure}[h!]
  \includegraphics[scale=0.42,natwidth=1546,natheight=1158]{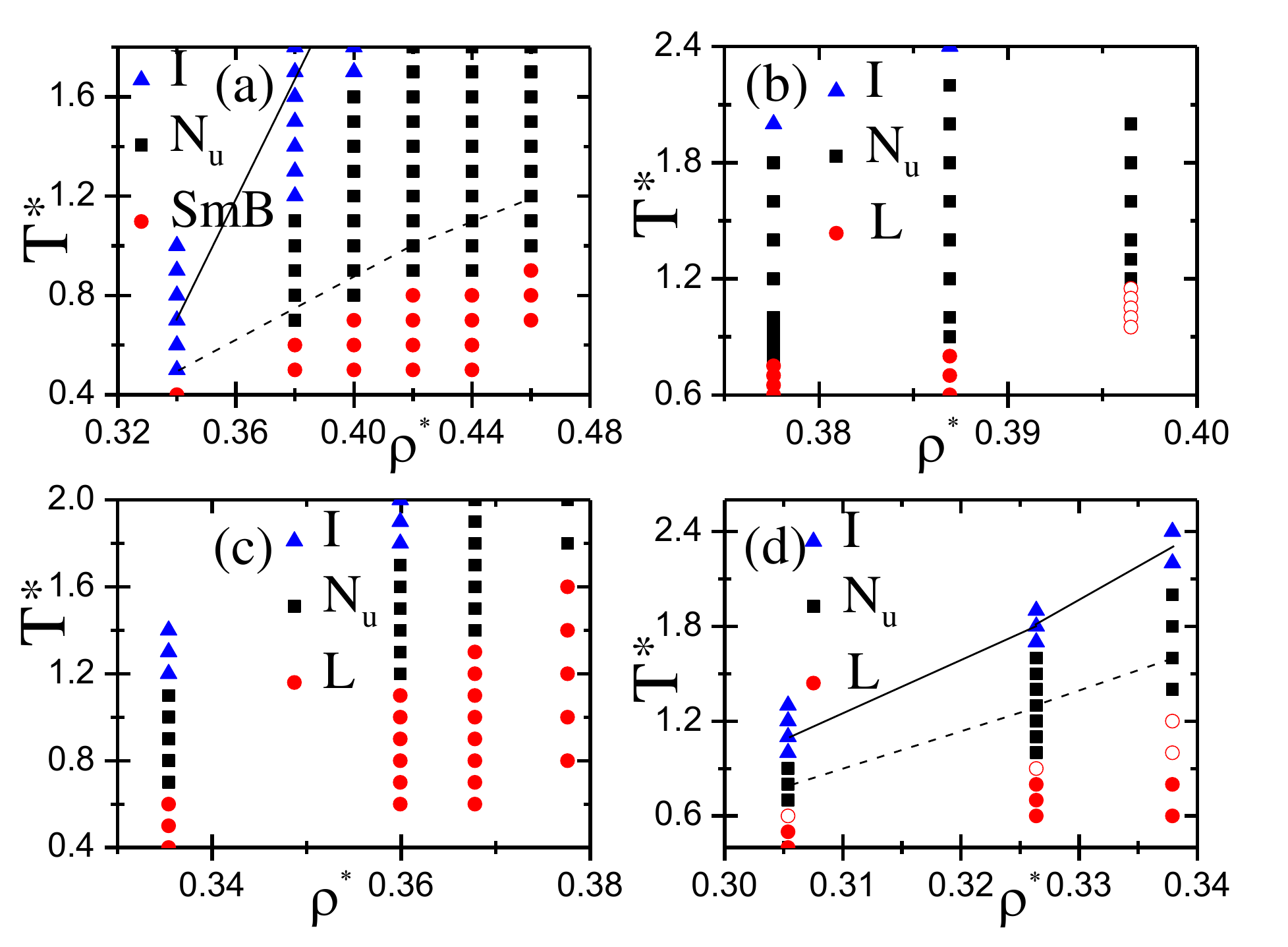}
  \caption{ (a) Tentative state diagram of a GB-SS mixture with (a)  $\sigma_{\text{s}}^*=1$ and $x_{\text{r}}=0.8$, (b)  $\sigma_{\text{s}}^*=1.5$ and $x_{\text{r}}=0.8$, (c)  $\sigma_{\text{s}}^*=1.7$ and $x_{\text{r}}=0.8$ and (d)  $\sigma_{\text{s}}^*=2$ and $x_{\text{r}}=0.9$. The points on the diagram indicate the pairs $(T^*,\rho^{*})$ for which the actual simulations were performed. The solid and dashed lines indicate state transformations of the GB-DSS (magnetic) system. Abbreviations: Fully miscible isotropic (I), uniaxial nematic $(\text{N}_{\text{u}})$, uniaxial smectic (SmB) and lamellar (L) phase.}
  \label{fig:01}
  \end{figure}
\end{center}

\subsubsection{GB-SS mixtures with $\sigma^{*}_{\text{s}}=1.0$. }

Initially, we examine GB-SS mixtures with $\sigma^{*}_{\text{s}}=1.0$. 
We have performed simulations for systems consisting of $N=720$ particles for $\textit{x}_{\text{r}}=0.8$. Larger systems of $N=2000$  particles have also been studied (for selected state points) to account for finite size effects. Either cooling series starting from a high temperature isotropic liquid or a nematic phase (prepared by melting a crystal state) are performed, at specific total number density $\rho^*$. In some cases, heating series from ordered phases were performed as well to check for the reversibility of the phase behaviour.

A calculated temperature-density $(T^*,\rho^{*})$  diagram is presented in  Fig.~\ref{fig:01}a where we use different symbols for the three distinct phases. Specifically, fully miscible phases of different order are exhibited: (i) an isotropic (I), (ii) an uniaxial nematic $(\text{N}_{\text{u}})$ and (iii) a smectic (SmB) phase.

\begin{center}
  \begin{figure}[h!]
  \includegraphics[scale=0.40,natwidth=1985,natheight=1229]{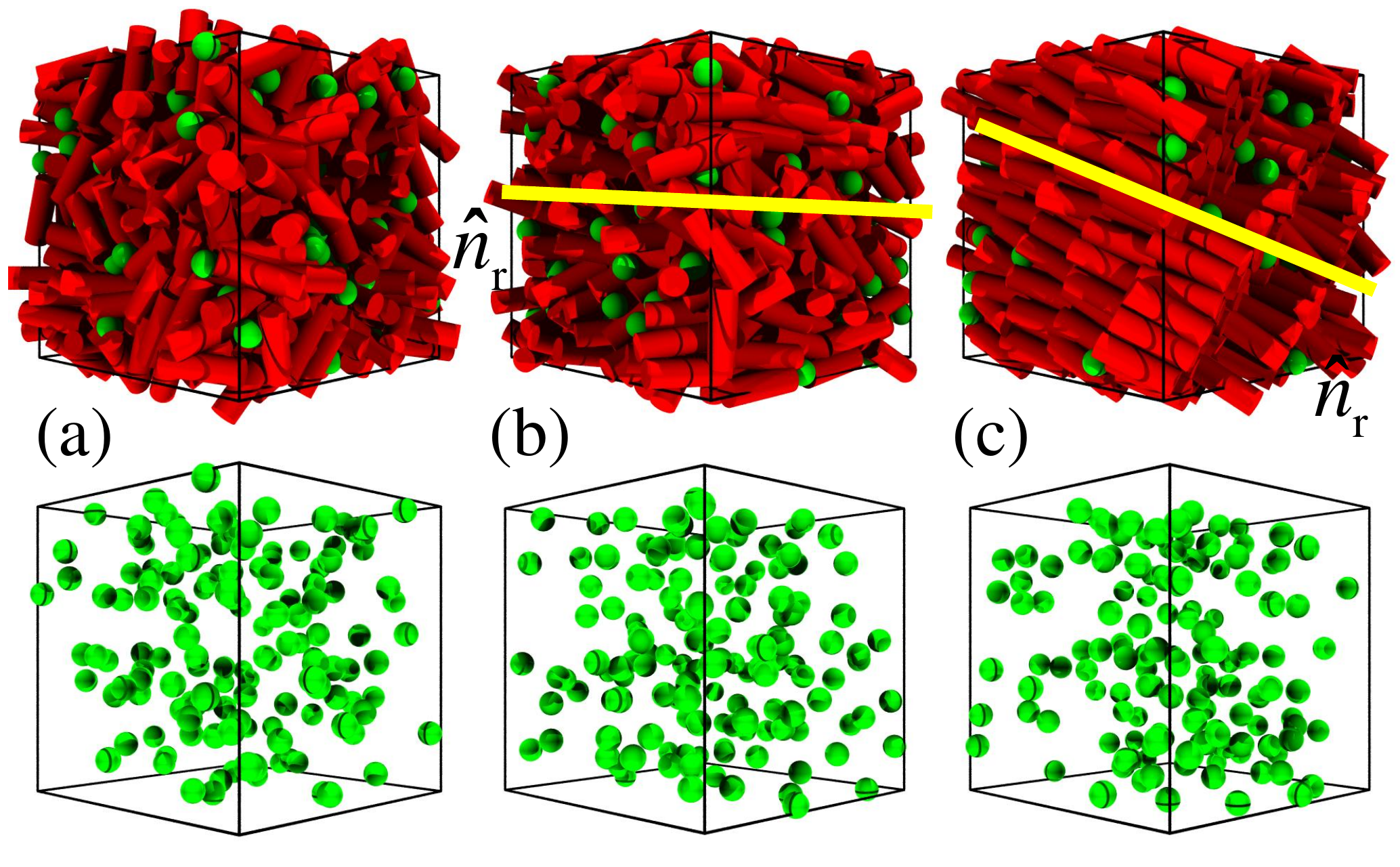}
  \caption{Representative snapshots of a GB-SS mixture with $\sigma_{\text{s}}^*=1$ and $x_{\text{r}}=0.8$ in various states: (a) isotropic state at $[(T^*,\rho^{*})=(1.8,0.40)]$, (b) uniaxial nematic ($\text{N}_{\text{u}}$) state at $[(T^*,\rho^{*})=(1.2,0.40)]$ and (c) smectic B (SmB) at $[(T^*,\rho^{*})=(0.8,0.40)]$. The direction of  $\hat{\mathbf{n}}_{\text{r}}$ is also shown. For clarity, the rod species have been removed from the simulation box in the bottom row.  }
  \label{fig:02}
  \end{figure}
\end{center}

\begin{center}
  \begin{figure}[h!]
  \includegraphics[scale=0.42,natwidth=1544,natheight=1221]{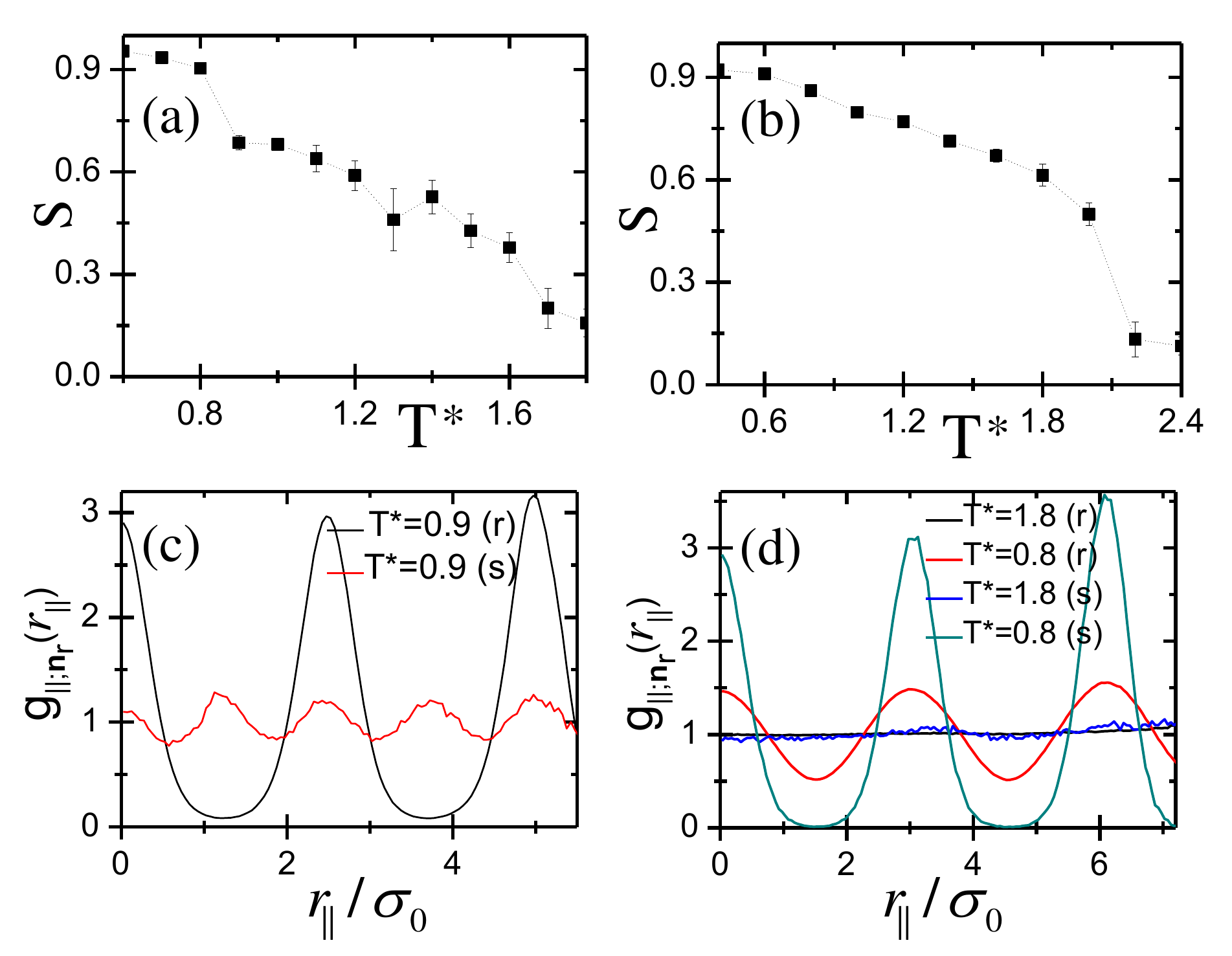}
  \caption{Order parameter as a function of temperature of a GB-SS mixture with (a) $\sigma^{*}_{\text{s}}=1.0$ at $\rho^{*}=0.40$ and (b) $\sigma^{*}_{\text{s}}=2.0$ at $\rho^{*}=0.338$. Pair correlation function $g_{\Vert;\hat{\mathbf{n}}_{\text{r}}}$ for the rod (r) and spherical (s) species with (c) $\sigma^{*}_{\text{s}}=1.0$ at $\rho^{*}=0.40$ and (d) $\sigma^{*}_{\text{s}}=2.0$ at $\rho^{*}=0.338$ for various $T^{*}$.} 
  \label{fig:03}
  \end{figure}
\end{center}

In the I phase both species lack long range orientational order. A characteristic snapshot of the isotropic phase is shown in Fig. \ref{fig:02}a. The orientational disorder of the rod species has been confirmed by calculating the order parameter $S^{(\text{r})}$. A representative order parameter vs temperature diagram for $\rho^*=0.4$ is given in Fig. \ref{fig:03}a. 
By performing a cooling series starting from an I-state  
the system first transforms into an uniaxial $\text{N}_{\text{u}}$ state 
where the rods possess a long range orientational order as can be seen from the increase of the order parameter (see also the snapshots in Fig. \ref{fig:02}b). In the  $\text{N}_{\text{u}}$ state both rods and spheres are homogeneously distributed and the phase has uniaxial symmetry with respect to the nematic director.  The pair correlation function $g_{\Vert;\hat{\mathbf{n}}_{\text{r}}}^{\left({\text{r}}\right)}\left(r_{\Vert}\right)$  confirms that both species are distributed homogeneously along $\hat{\mathbf{n}}_{\text{r}}$. 

By further decreasing the temperature, a state with translational order is obtained in which the rods are organized in periodic layers (see Fig \ref{fig:02}c).  
This behavior is also confirmed by the strong modulations in $g_{\Vert;\hat{\mathbf{n}}_{\text{r}}}^{\left({\text{r}}\right)}\left(r_{\Vert}\right)$ (see Fig \ref{fig:03}c). The rods possess local hexagonal order within the layers similar to monodispersed (GB) rod systems \cite{Miguel99}. Interestingly, in this SmB phase there is a tendency of the spherical particles to organize into linear arrangements along the director of the phase as it can be seen from inspecting the snapshots shown in Fig \ref{fig:02}c. This is also reflected by the behavior of the function $g_{\Vert;\hat{\mathbf{n}}_{\text{r}}}^{\left({\text{s}}\right)}\left(r_{\Vert}\right)$ which displays modulations within the SmB phase with a periodicity of approximately one $\sigma_0$ (see Fig. \ref{fig:03}c). This implies that the spheres are incorporated within the layers of rods rather than positioning between successive layers of rods through microphase separation. We discuss this issue in more detail in the next subsection.

\subsubsection{GB-SS mixtures with $1.5\leq \sigma^{*}_{\text{s}}\leq 2.0$. }

\begin{center}
  \begin{figure}[h!]
  \includegraphics[scale=0.40,natwidth=1481,natheight=896]{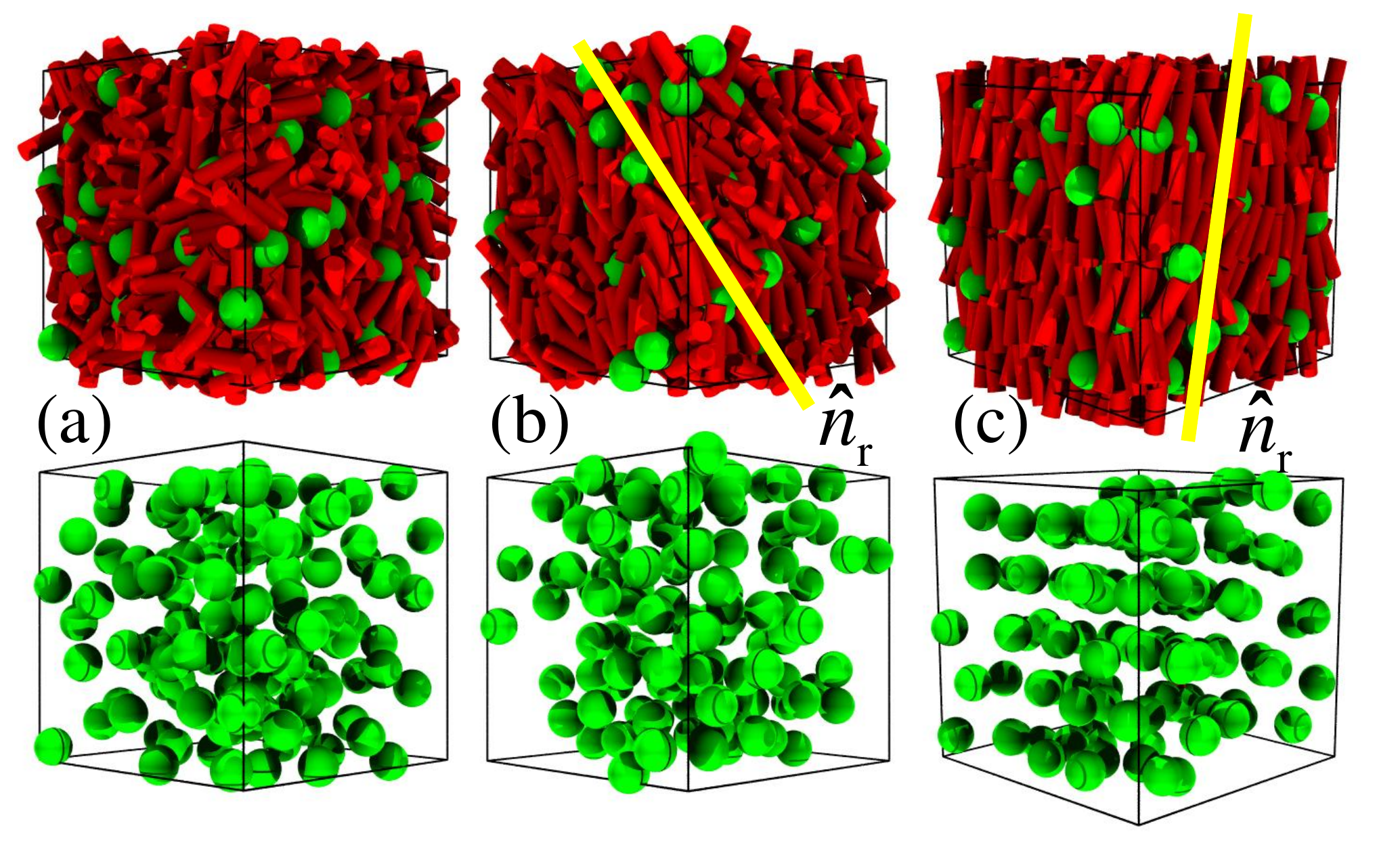}
  \caption{Representative snapshots of a GB-SS mixture with $\sigma_{\text{s}}^*=2$,  for $x_{\text{r}}=0.9$ in various states: (a) isotropic state  at $[(T^*,\rho^{*})=(2.4,0.338)]$, (b) uniaxial nematic ($\text{N}_{\text{u}}$) state at $[(T^*,\rho^{*})=(1.8,0.338)]$ and (c) Lamellar (L) state at $[(T^*,\rho^{*})=(0.8,0.338)]$. For further details see Fig. 2.  }
  \label{fig:04}
  \end{figure}
\end{center}

We now turn to study GB-SS mixtures with larger diameter of $\sigma^{*}_{\text{s}} \geq 1.5 $. Specifically, we have performed simulations for systems of $N=720$ particles with $\sigma^{*}_{\text{s}}=1.5$  and $\sigma^{*}_{\text{s}}=1.7$  for  $x_{\text{r}}=0.8$ and of $N=1251$ particles with $\sigma^{*}_{\text{s}}= 2.0$ for $x_{\text{r}}=0.9$. Temperature vs density  $(T^*,\rho^{*})$  diagrams are shown in Fig. \ref{fig:01}b-d. A common feature with the previous system ($\sigma^{*}_{\text{s}}= 1.0$) is that fully miscible I and $\text{N}_{\text{u}}$ phases are obtained. Both I and $\text{N}_{\text{u}}$ lack long range positional order. Typical snapshots of these phases are shown in Fig \ref{fig:04}a-b. An example that monitors the order parameter as a function of the temperature within the $\text{N}_{\text{u}}$ state is given in Fig. \ref{fig:03}b for a system with $\sigma^{*}_{\text{s}}= 2$. 

Remarkably, the layered phase that is obtained for lower temperatures is not a common smectic phase. 
The translational ordered phase consists of alternating rod-rich and sphere-rich layers indicating 
spontaneous microphase separation (see Fig. \ref{fig:04}c). Particularly noteworthy is that both $g_{\Vert;\hat{\mathbf{n}}_{\text{r}}}^{\left({\text{r}}\right)}\left(r_{\Vert}\right)$  and  $g_{\Vert;\hat{\mathbf{n}}_{\text{r}}}^{\left({\text{s}}\right)}\left(r_{\Vert}\right)$ functions are  modulated in the smectic phase revealing translational ordering along the director $\hat{\mathbf{n}}_{\text{r}}$.  It should be noted that there is also considerable interdigitation especially for the rodlike particles. This type of smectic phase is termed lamellar (L) and has already been observed both experimentally \cite{Fraden98} and theoretically\cite{PeroukidisMix,Cinacchi04,Dogic2000}. In particular, implementations of MC simulations  \citep{Dogic2000} and Onsager type theory \citep{PeroukidisMix,Cinacchi04} for purely repulsive (hard-core) rod-sphere mixtures have shown that the lamellar state is thermodynamically stable for  $\sigma^{*}_{\text{s}}=1.0$. Interestingly, 
we do not observe lamellar organization for the GB-SS mixture with $\sigma^{*}_{\text{s}}=1.0$. The reason for this is because, in our model, the side-by-side rod-sphere configuration is energetically favourable. Therefore, three spheres can fit into one layer spacing of rods without destroying the order of the SmB state. It should be noted that the side-by-side configuration is not favoured for rod-sphere mixtures interacting through hard pair potentials (such as hard spherocylinders and hard spheres\cite{Dogic2000}). In this case, in order to reduce the excluded volume of spherocylinder-sphere pairs the species microphase separate and form a lamellar phase\cite{Dogic2000}.  For systems with $\sigma^{*}_{\text{s}}\geq 1.5$ a side-by-side rod-sphere configuration is also the most energetically favourable one; in this case, though, this configuration destroys the smectic order of the rods since the spheres do not fit into one layer spacing of the rods. In order to preserve a layered structure at lower temperatures the system microphase separates into a lamellar phase with a high degree of interdigitation for the rod species.

\subsection{Binary mixtures of rods and dipolar soft  spheres (GB-DSS)}
In this section we examine the role of the dipolar interactions on the structure of binary GB-DSS mixtures. The configurations of the non-magnetic reference systems (see section 3.1) are used as initial configurations for the GB-DSS systems. We have considered spheres with a relatively large central permanent point dipole moment of strength $m^*=m/ \sqrt{\epsilon_0\sigma_{\text{s}}^3}=3 $. Inspired by experimental values \cite{Klokkenburg06,Odenbach2014,Ilg13} the coupling parameter $\lambda= m^2/\textit{k}_{B}\textit{T}\sigma_{\text{s}}^3 $ takes values greater than $3.5$.

\subsubsection{GB-DSS mixture with $\sigma^{*}_{\text{s}}=1.0$. } Initially we present a brief summary of the behavior of a GB-DSS mixture where the diameter of the spheres is equal to the width of the rods. A detailed analysis is presented elsewhere \cite{StavarXiv}. The topology of the $(T^*-\rho^*)$ diagram is similar to the respective GB-SS system that is shown in Fig \ref{fig:01}a; nevertheless, the morphology of the DSS particles within the phases is completely different. The solid and dashed lines in Fig \ref{fig:01}a indicate I-uniaxial nematic $\text{N}_{\text{u}}$ and $\text{N}_{\text{u}}$-uniaxial smectic SmB transformations, respectively. An important finding is that the I-state is destabilized in favor of the LC phases in the absence of any external stimuli. In the I-state, the DSS particles self-assemble into isotropic networks of wormlike chains. Remarkably, in the $\text{N}_{\text{u}}$ state the ferromagnetic chains spontaneously align along the LC director  $\boldsymbol{\hat{n}}_{\text{r}}$. Overall, the ferromagnetic chains are randomly arranged ``up'' or ``down'' and the system does not exhibit a net magnetization. Finally, within the smectic state the rod particles form a SmB phase with well defined layers; the ferromagnetic chains penetrate and pass through these layers. This result is in accordance with the observed configurations of the corresponding  GB-SS mixtures in which the SS particles fit into the smectic layers (see Sec.~3.1.1).

\subsubsection{GB-DSS mixture with $\sigma^{*}_{\text{s}}=1.5$ }
\begin{center}
  \begin{figure}[h!]
  \includegraphics[scale=0.41,natwidth=1552,natheight=632]{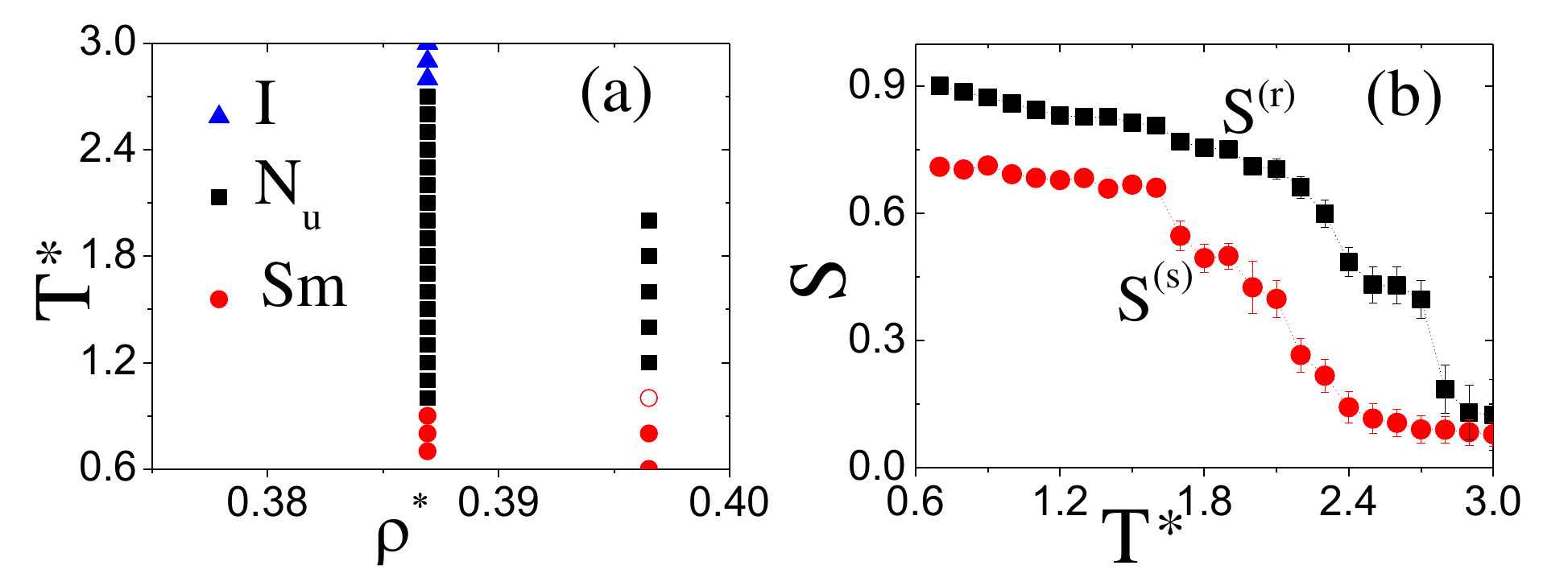}
  \caption{ (a) State diagram of a GB-DSS mixture with $\sigma_{\text{s}}^*=1.5$ $(x_{\text{r}}=0.8)$, involving isotropic (I), uniaxial nematic ($\text{N}_{\text{u}}$) and uniaxial interdigitated smectic (Sm). The symbols indicate state points where the actual simulations were performed. (b) Global order parameters as functions of  $T^*$ at $\rho^*=0.387$. }
  \label{fig:05}
  \end{figure}
\end{center}

From here on we examine the role of the rod-sphere size ratio on the self-organization of GB-DSS mixtures. We address basic questions regarding (i) the morphology of the DSS particles inside the LC matrix and (ii) how the DSS ordering influences the LC matrix and vice versa. As we have seen in Sec.~3.1,  the translational order changes considerably by tuning the rod-sphere ratio already for the case of the (non-magnetic) GB-SS mixture. 
First, we increase the size of the dipolar spheres to $\sigma_{\text{s}}^*=1.5$. 
A temperature-density $(T^*-\rho^*)$ diagram has been calculated for binary mixtures of $N=720$ particles and $x_{\text{r}}=0.8$ (see Fig \ref{fig:05}a). This corresponds to volume fraction ratios (rods to spheres) $\phi_{\text{r}}/\phi_{\text{s}}\cong 3.54 $ which is of the same order of magnitude that can be achieved in real colloidal suspensions \cite{privcomm}. Isotropic (I), uniaxial nematic $\text{N}_{\text{u}}$ and highly interdigitated smectic (Sm) phases are obtained.

\begin{center}
  \begin{figure}[h!]
  \includegraphics[scale=0.40,natwidth=1521,natheight=959]{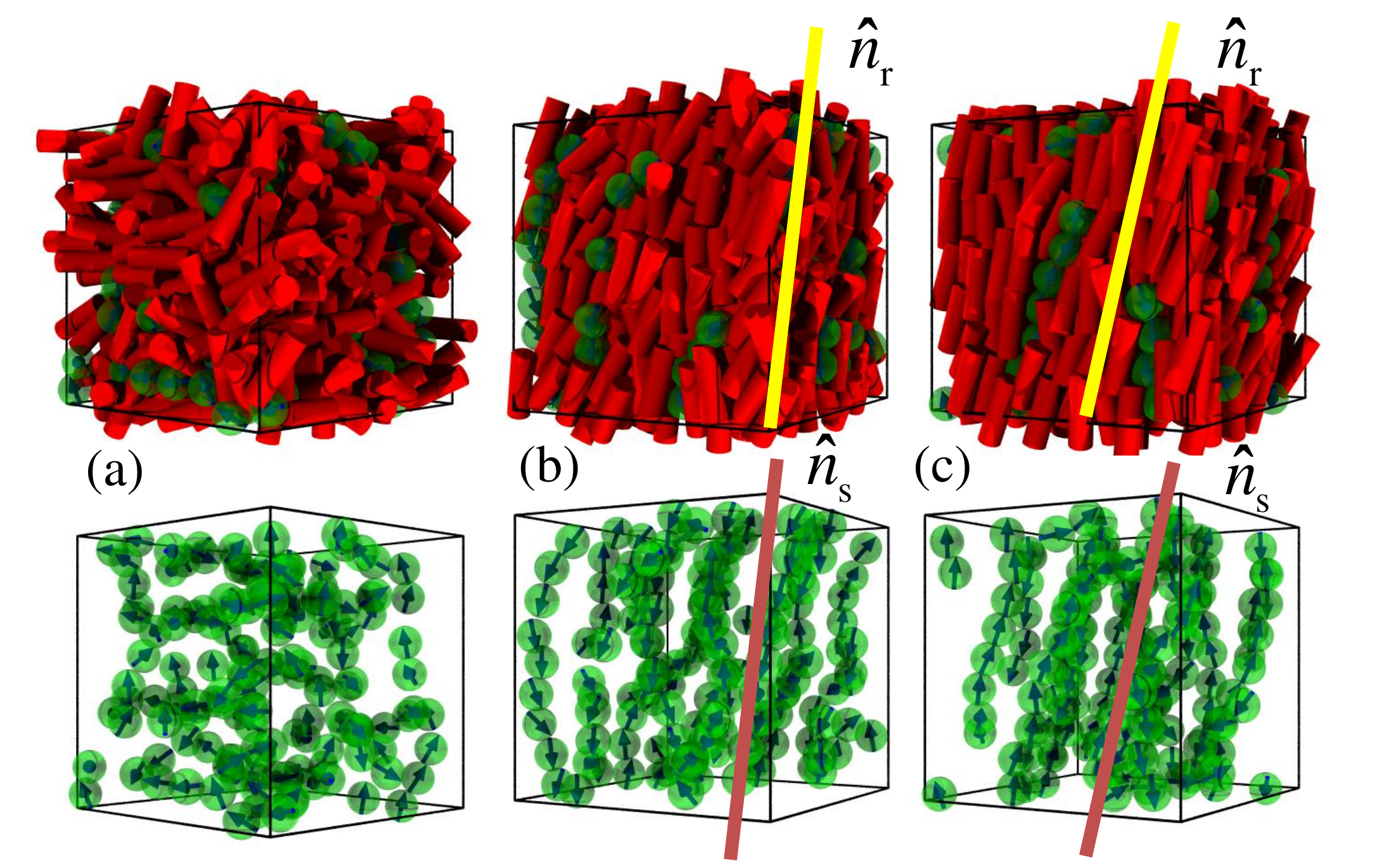}
  \caption{Representative simulation snapshots of a GB-DSS mixture with $\sigma_{\text{s}}^*=1.5$ for $x_{\text{r}}=0.8$ in various states: (a) isotropic state at $[(T^*,\rho^{*})=(2.8,0.387)]$ (b) uniaxial nematic ($\text{N}_{\text{u}}$) state at $[(T^*,\rho^{*})=(1.6,0.387)]$, and (c) highly interdigitated smectic (Sm) state at $[(T^*,\rho^{*})=(0.8,0.387)]$. For clarity, only DSS particles are shown in the bottom row. The directions $\hat{\mathbf{n}}_{\text{r}}$ and $\hat{\mathbf{n}}_{\text{s}}$ are indicated by thick lines.  }
  \label{fig:06}
  \end{figure}
\end{center}

\begin{center}
  \begin{figure}[h!]
  \includegraphics[scale=0.40,natwidth=1575,natheight=1260]{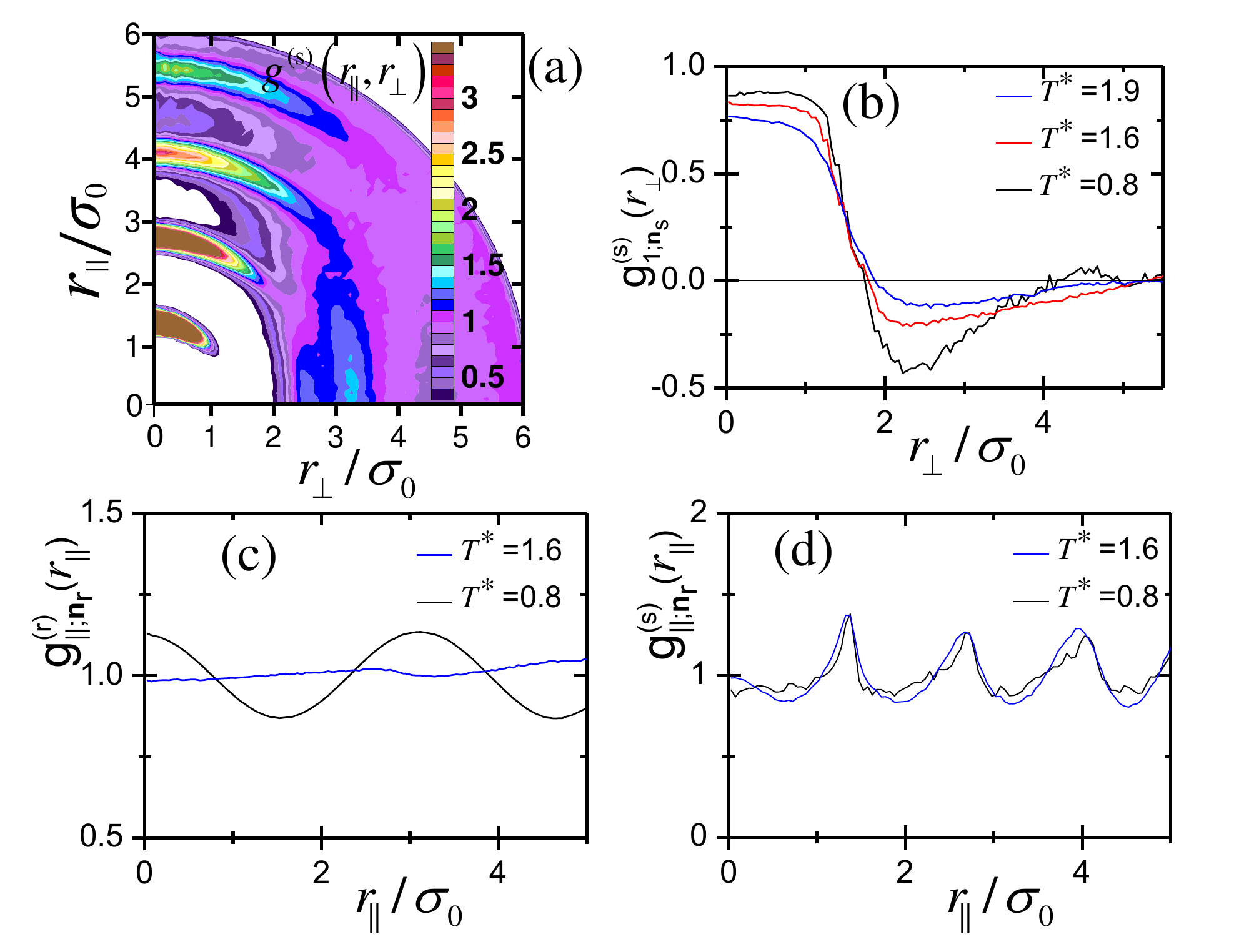}
  \caption{Representative correlation functions for a GB-DSS mixture with $\sigma_{\text{s}}^*=1.5$ and $x_{\text{r}}=0.8$: (a) two-dimensional pair correlation function $g^{\left({\text{s}}\right)}\left(r_{\Vert},r_{\bot}\right)$  in the uniaxial nematic  ($\text{N}_{\text{u}}$) state at $[(T^*,\rho^{*})=(1.6,0.387)]$. Distribution functions (b) $g_{1;\hat{\mathbf{n}}_{\text{s}}}^{\left({\text{s}}\right)}\left(r_{\bot}\right)$, (c) $g_{\Vert;\hat{\mathbf{n}}_{\text{r}}}^{\left({\text{r}}\right)}\left(r_{\Vert}\right)$ and (d) $g_{\Vert;\hat{\mathbf{n}}_{\text{r}}}^{\left({\text{s}}\right)}\left(r_{\Vert}\right)$ for $\rho^{*}=0.387$ and various $T^{*}$.  }
  \label{fig:07}
  \end{figure}
\end{center}

In the I state both species do not possess long range orientational and positional order. The DSS particles self-assemble into ferromagnetic chains. A representative simulation snapshot is shown in Fig \ref{fig:06}a. A similar structure is also found in monodispersed dilute DSS systems \cite{Klapp05,Jordan2011}. By reducing the temperature starting from the isotropic phase the system undergoes an I-$\text{N}_{\text{u}}$ phase transformation. A significant increment of the order parameter $S^{({\text{a}})}$ of both species is clearly demonstrated in a $(S^{(\text{a})}-T^*)$ diagram for $\rho^*=0.387$ that is shown in Fig \ref{fig:05}b.  Remarkably, the ferromagnetic chains are spontaneously unwrapped within the nematic phase forming ferromagnetic  chains that are on average parallel to the director $\hat{\mathbf{n}}_{\text{r}}$. Interestingly, there is a notable delay concerning the magnitude of the order parameter $S^{(\text{s})}$ in comparison to $S^{(\text{r})}$ that jumps to 0.4 at the I-$\text{N}_{\text{u}}$ transformation. This signifies a lower alignment of the ferromagnetic chains by the LC matrix in comparison to the GB-DSS mixture with $\sigma^{*}_{\text{s}}=1.0$ (cf. Fig.~1b in Ref.\citenum{StavarXiv}).  The corresponding value of $B\simeq 1.0$ means that the directors of the species are, on average, parallel to each other rendering the phase  uniaxial. A characteristic snapshot of the $\text{N}_{\text{u}}$ phase is shown in Fig \ref{fig:06}b. A  signature of the arrangement of the DSS particles are the periodic arcs that appear in the anisotropic  $g^{\left({\text{s}}\right)}\left(r_{\Vert},r_{\bot}\right)$  pair correlation function (parallel to the dipole in comparison to the perpendicular direction) (see Fig \ref{fig:07}a).

Interestingly, the nematic phase is significantly enhanced in favor of the isotropic phase in comparison to the system without dipolar interactions (see Fig.~\ref{fig:01}b and Fig.~\ref{fig:05}a). Therefore, there is an interplay in which the LC rods induce orientational order to ferromagnetic chains and vice versa. Even though the chains are polar, the phase does not show spontaneous magnetization since the polar chains are arranged into an antiparallel manner. This has been confirmed by the polar order parameter $P_{1}$ that takes a value near zero. The presence of polar domains beyond a single chain is also excluded, since the $g_{1;\hat{\mathbf{n}}_{\text{s}}}\left(r_{\bot}\right)$ function takes large positive values near the origin and vanishes already at small distances (see Fig \ref{fig:07}b). It is important to note that monodispersed systems of dipolar spheres \cite{Klapp05,Jordan2011}, in the absence of external fields, do not exhibit spontaneous orientational order for the densities considered here. The homogeneous positional distribution of rods along the director is confirmed by the structureless function $g_{\Vert;\hat{\mathbf{n}}_{\text{r}}}^{\left({\text{r}}\right)}\left(r_{\Vert}\right)$ (see Fig \ref{fig:07}c). The light modulations of $g_{\Vert;\hat{\mathbf{n}}_{\text{r}}}^{\left({\text{s}}\right)}\left(r_{\Vert}\right)$ in the $\text{N}_{\text{u}}$ state with periodicity of approximately one molecular diameter shown in Fig.~\ref{fig:07}d occurs due to the correlations of spheres that belong to the same chain; furthermore, the absence of a peak at the origin indicates the formation of chains that slide along the director direction, thus preventing the development of strong positional correlations between them. The morphologies found within the nematic phase in the system studied here are similar to the one observed for the smaller DSS (see section 3.2.1.). 

At lower temperatures, the system forms a layered structure with highly interdigitated rods (see the snapshots in Fig.~\ref{fig:06}c), also reflected by the light modulations of $g_{\Vert;\boldsymbol{\hat{n}}_{\text{r}}}^{\left({\text{r}}\right)}\left(r_{\Vert}\right)$ (see the black curve in Fig.~\ref{fig:07}c). This is in contrast to the system with smaller spheres ($\sigma^{*}_{\text{s}}=1.0$) in which well defined layers are formed. Consequently, an increment of the diameter of the DSS strongly disturbs the translational order. An important finding is that the magnetic chains penetrate the smectic layers rather than lying between the layers of rods (as it occurs in the corresponding GB-SS mixture forming an uniaxial smectic phase). The self-assembly of the DSS into magnetic chains alters dramatically the translational distribution of the DSS within the liquid crystalline matrix. It should be noted that the lamellar order is destroyed even when a lamellar configuration for the GB-SS mixture is being used as initial configuration. Hence, a parallel arrangement of chains and rods is preferred instead of a perpendicular one. Furthermore, the inplane translational order of the rods indicates an isotropic positional arrangement characteristic of a SmA phase.

\subsubsection{GB-DSS mixture with $\sigma^{*}_{\text{s}}=1.7$ }
We have further increased the diameter of the DSS particles to $\sigma^{*}_{\text{s}}=1.7$ and have studied the morphologies of systems consisting of $N=720$ particles at $x_{\text{r}}=0.8$. Initially, we performed a cooling series starting from an isotropic phase. The system undergoes phase transformations to an uniaxial nematic state and, at lower temperatures, to a "glassy" state that does not possess translational long-range order. This is a firm indication that the DSS particles exert here a stronger perturbation on the LC matrix than for systems with smaller DSS sizes $\sigma^{*}_{\text{s}} \leq 1.5$. It should also be noted that the  volume fraction ratio $\phi_{\text{r}}/\phi_{\text{s}}\cong 2.44 $ is smaller than in the systems studied in previous subsections.

\begin{center}
  \begin{figure}[h!]
  \includegraphics[scale=0.42,natwidth=1591,natheight=615]{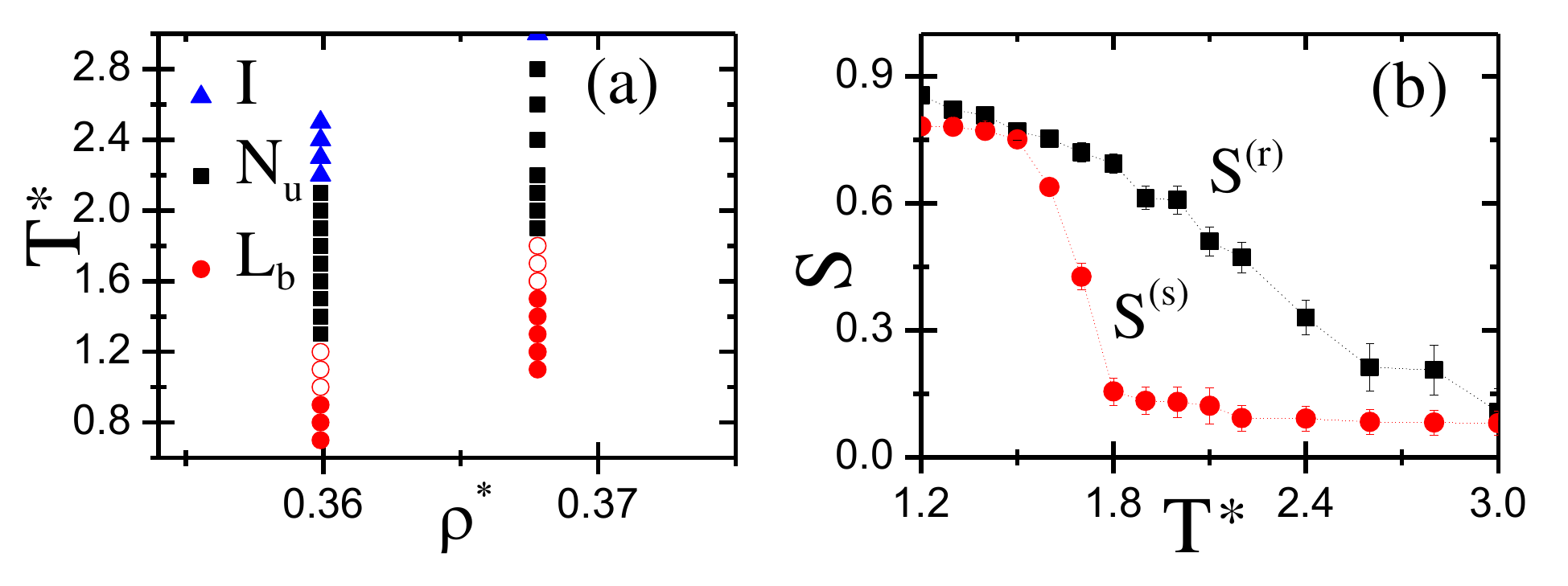}
  \caption{ (a) State diagram of a GB-DSS mixture with $\sigma_{\text{s}}^*=1.7$ $(x_{\text{r}}=0.8)$, involving isotropic (I), nematic $(\text{N}_{\text{u}})$ and biaxial lamellar $(\text{L}_{\text{b}}) $. The symbols indicate state points where the actual simulations were performed. (b) Global order parameters as functions of  $T^*$ at $\rho^*=0.368$. }
  \label{fig:08}
  \end{figure}
\end{center}

\begin{center}
  \begin{figure}[h!]
  \includegraphics[scale=0.44,natwidth=1249,natheight=1034]{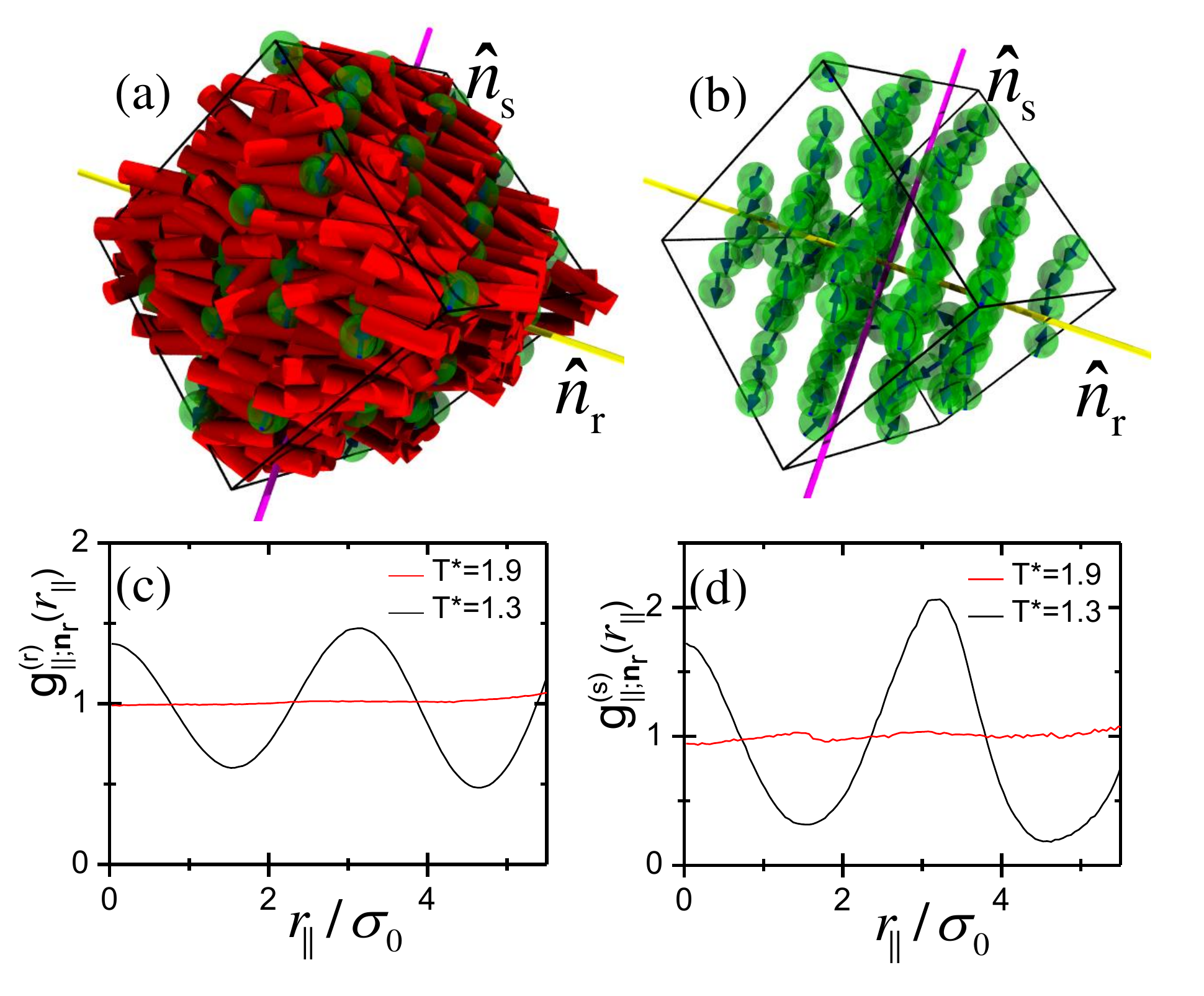}
  \caption{Representative simulation snapshots for a GB-DSS mixture with $\sigma_{\text{s}}^*=1.7$,  for $x_{\text{r}}=0.8$ in the biaxial lamellar $(\text{L}_{\text{b}})$ state at $[(T^*,\rho^{*})=(1.3,0.387)]$ (a) both species are shown and (b) DSS particles are shown for clarity.  The directions $\hat{\mathbf{n}}_{\text{r}}$ and $\hat{\mathbf{n}}_{\text{s}}$ are indicated by thick lines. Distribution functions:  (c) $g_{\Vert;\hat{\mathbf{n}}_{\text{r}}}^{\left({\text{r}}\right)}\left(r_{\Vert}\right)$ and (d) $g_{\Vert;\hat{\mathbf{n}}_{\text{r}}}^{\left({\text{s}}\right)}\left(r_{\Vert}\right)$ for $\rho^{*}=0.387$ and various $T^{*}$.  }
  \label{fig:9}
  \end{figure}
\end{center}

To get a more detailled insight into the behavior of the system we have also performed a heating series using lamellar structures obtained from the corresponding GB-SS mixture as initial configurations. A $(T^{*},{\rho}^{*})$ diagram obtained for two different densities is shown in Fig \ref{fig:08}a. 
For clarity we describe the system at $\rho^{*}=0.368$. Interestingly, the lamellar structure persists also for the GB-DSS mixture. A visual inspection of representative snapshots (see Fig \ref{fig:9}a-b) reveals a microseparation into alternating rod-rich and DSS-rich regions. Furthermore, both species exhibit significant orientational order as reflected by increased order parameter values $S^{({\text{a}})} \geq 0.7$ (see Fig \ref{fig:08}b).  The modulations  of the  $g_{\Vert;\hat{\mathbf{n}}_{\text{r}}}^{\left({\text{a}}\right)}\left(r_{\Vert}\right)$ function clearly indicate the formation of layers (see Fig \ref{fig:9}c-d). A pronounced interdigitation of the rod species is also seen from the relatively slight modulation of $g_{\Vert;\hat{\mathbf{n}}_{\text{r}}}^{\left({\text{r}}\right)}\left(r_{\Vert}\right)$ in Fig \ref{fig:9}c. The DSS self-assemble into ferromagnetic chains. As these are arranged in an antiparallel manner no net magnetization is exhibited. The most striking result is that the director of the rods is perpendicular to the director of the DSS with $B \simeq 0$. We term this state biaxial lamellar  $L_{\text{b}} $.

Upon further heating, the system undergoes a transformation from a lamellar phase to an uniaxial nematic phase with respect to the rod species. Similar to the system with $\sigma^{*}_{\text{s}}=1.5$ there is a notable delay of the magnitude of the $S^{(\text{s})}$ order parameter in comparison to $S^{(\text{r})}$ at the transition to the nematic state (see Fig \ref{fig:08}b). 
In summary, the above results indicate that (i) larger DSS particles destabilize conventional smectic order (either SmA or SmB) and (ii)  the perturbation of the DSS particles to the LC host is such that the LC matrix does not induce, within the $\text{N}_\text{u}$ state, any significant alignment to the ferromagnetic chains either perpendicular or parallel to the mean orientation of the rods.

\subsubsection{GB-DSS mixture with $\sigma^{*}_{\text{s}}=2.0$ } Here, we describe briefly the structure of a GB-DSS mixture with relatively large diameter, $\sigma^{*}_{\text{s}}=2.0$. For a detailed consideration of this system see Ref.\citenum{StavarXiv}. The self-organization of the magnetic particles in the nematic state differs in comparison to the mixture with $\sigma^{*}_{\text{s}} \leq 1.7$. Remarkably, by decreasing the temperature starting from an isotropic state these systems exhibit spontaneously biaxial nematic ordering in which, on average, the director $\hat{\mathbf{n}}_{\text{r}}$ of the rods is perpendicular to the director $\hat{\mathbf{n}}_{\text{s}}$ of the DSS particles. Upon further cooling, a biaxial lamellar $\text{L}_{\text{b}} $ is found similar to the GB-DSS mixture with $\sigma^{*}_{\text{s}}=1.7$. In the $\text{L}_{\text{b}} $-state the ferromagnetic chains align, on average,  perpendicularly to the director $\hat{\mathbf{n}}_{\text{r}} $, and are arranged into an antiparallel manner resulting no net magnetization. 

\subsection{Response to external magnetic fields}

\subsubsection{GB-DSS mixture subject to external field with $\sigma^{*}_{\text{s}}=1.0$. }

\begin{center}
  \begin{figure}[h!]
  \includegraphics[scale=0.44,natwidth=1617,natheight=559]{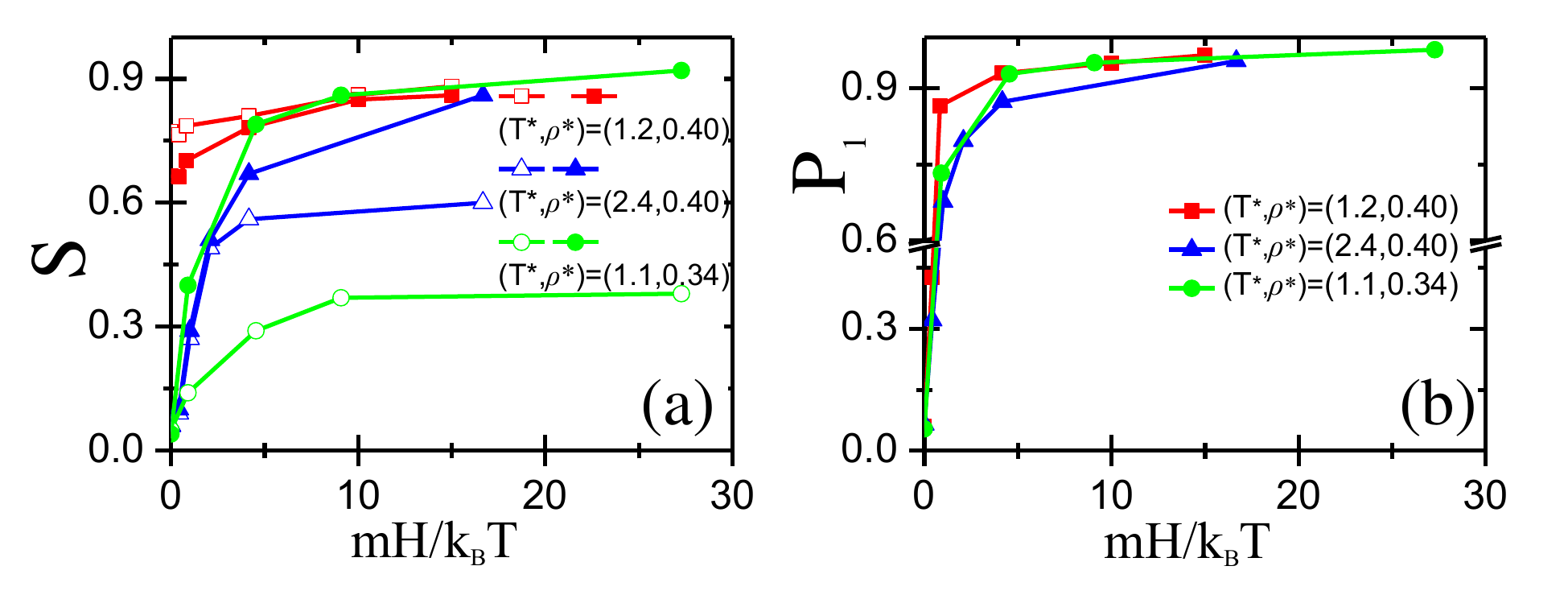}
  \caption{ (a) Nematic Order parameter and (b) polar order parameter as a function of the external magnetic field strength for a GB-DSS mixture with $\sigma^{*}_{\text{s}}=1.0$. Open symbols correspond to rods and solid symbols to dipolar spheres.}
  \label{fig:10}
  \end{figure}
\end{center}

In this section we examine the response of a GB-DSS mixture to an homogeneous magnetic field given by $\mathbf{H}=H\hat{\mathbf{z}}$ at various field strengths $H^{*}=m H/k_{B}T$. The field $\mathbf{H}$ is coupled to the permanent dipole $\mathbf{m}_{i}$ of particle $i$ through the potential $U_{i}=-\mathbf{\hat{m}}_{i} \cdot \mathbf{H}$. Three representative state points $[(T^*,\rho^{*})=(1.1,0.34)]$, $[(T^*,\rho^{*})=(2.4,0.40)]$ and  $[(T^*,\rho^{*})=(1.2,0.40)]$ have been examined; the first two correspond to the isotropic and the third to the nematic phase of the field-free GB-DSS mixture (see the boundary lines in Fig \ref{fig:01}a). The nematic and polar order parameters are shown in Fig~\ref{fig:10} as a function of the field strength. The director $\hat{\mathbf{n}}_\text{s}$ of the DSS particles is, on average, parallel to $\mathbf{H}$. Therefore, the polar order parameter defined in Eq.~(\ref{eq.polar}) monitors the total magnetization of the system (i.e. the sum over the normalized magnetic moments). When the field is off in the isotropic state, for  $[(T^*,\rho^{*})=(1.1,0.34)]$ and  $[(T^*,\rho^{*})=(2.4,0.40)]$, the polar and the nematic order parameters are nearly zero. The nematic order parameter values for both the rod and the DSS species increase by increasing the strength of the magnetic field until saturation is reached at 
field strength $H^*\simeq 15$. Therefore, the rods species exhibit a field induced I-$\text{N}_{\text{u}}$ transformation. The DSS particles form ferromagnetic chains that are oriented parallel to the direction of the magnetic field giving rise to a net polarity that also saturates for even small magnetic fields (see Fig.~\ref{fig:10}). Remarkably, the linear ferromagnetic chains induce orientational order to the rodlike particles even for the isotropic state. The calculated nematic directors of the species are on average parallel to each other and an uniaxial nematic phase is exhibited. Notably, the reverse phenomenon (in comparison to the systems described in section 3.2) is obtained here which has already been observed for real colloidal suspensions of rodlike and magnetic particles \cite{Kredentser13}: Under an external homogeneous field the magnetic spheres self assemble and form rodlike entities that induce nematic order to colloidal rods even in the isotropic phase. The most interesting finding is that the saturation value of the nematic order parameter of the rod species is larger for higher densities ($\rho^{*}=0.40$) than for lower 
densities ($\rho^{*}=0.34$). This means that the optical properties (such as birefringence in real systems) are expected to depend sensitively on the volume fraction of rods. Preliminary experimental results \cite{privcomm} for colloidal suspensions of rodlike particles with magnetic spheres indicate that the saturation value of birefringence increases by an increment of the volume fraction of rods at constant volume fraction of spheres. Finally, for the nematic state $[(T^*,\rho^{*})=(1.2,0.40)]$ the nematic order parameter is slightly increased under an external field (see Fig \ref{fig:10}a). The polarity, on the other hand, increases significantly by changing from nearly zero to over 0.9 (see Fig \ref{fig:10}b).

\subsubsection{GB-DSS mixture subject to external field with $\sigma^{*}_{\text{s}}=2.0$. }

\begin{center}
  \begin{figure}[h!]
  \includegraphics[scale=0.44,natwidth=1563,natheight=598]{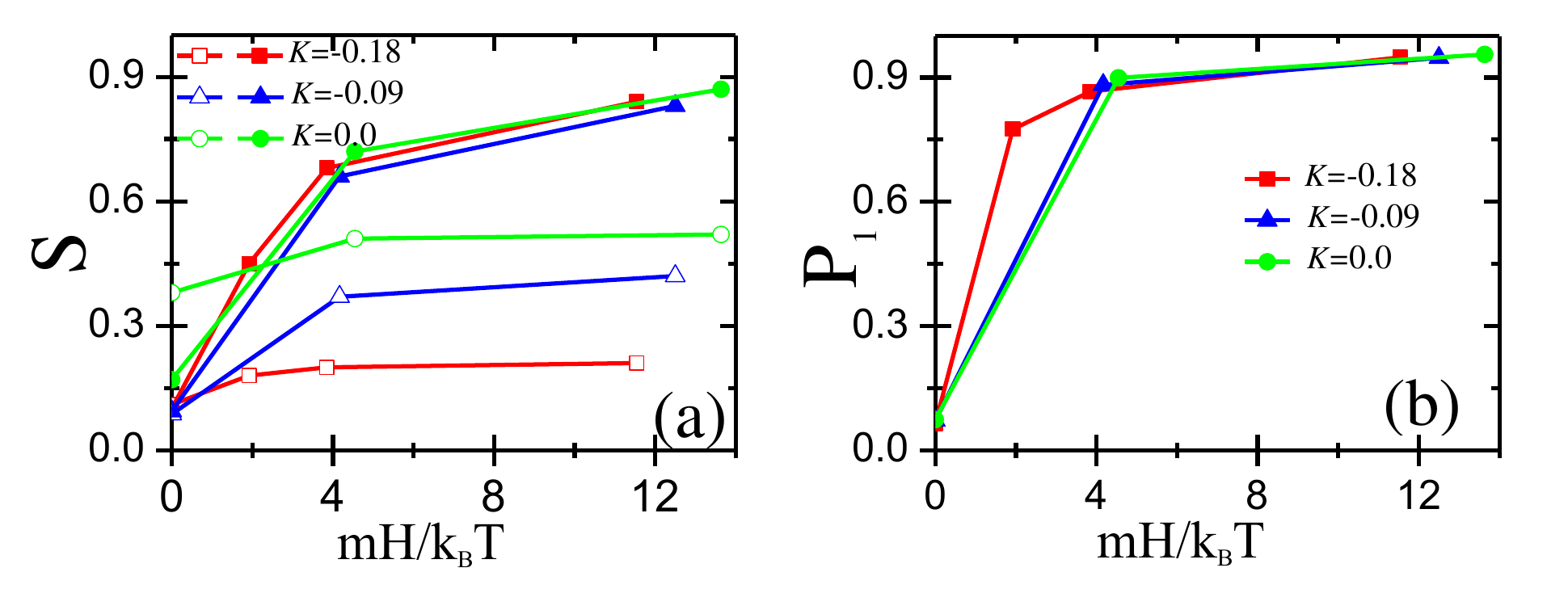}
  \caption{ (a) Nematic Order parameter and (b) polar order parameter as a function of the external magnetic field strength for a GB-DSS mixture with $\sigma^{*}_{\text{s}}=2.0$. We use the parameter $K$ defined as $K=1-T^{*}/T^{*}_{\text{I}\text{N}_\text{u}}$ to indicate how far the system is from the I-$\text{N}_\text{u}$ transformation temperature. For further details see the manuscript. Open symbols correspond to rods and solid symbols to spheres.}
  \label{fig:11}
  \end{figure}
\end{center}

\begin{center}
  \begin{figure}[h!]
  \includegraphics[scale=0.75,natwidth=890,natheight=780]{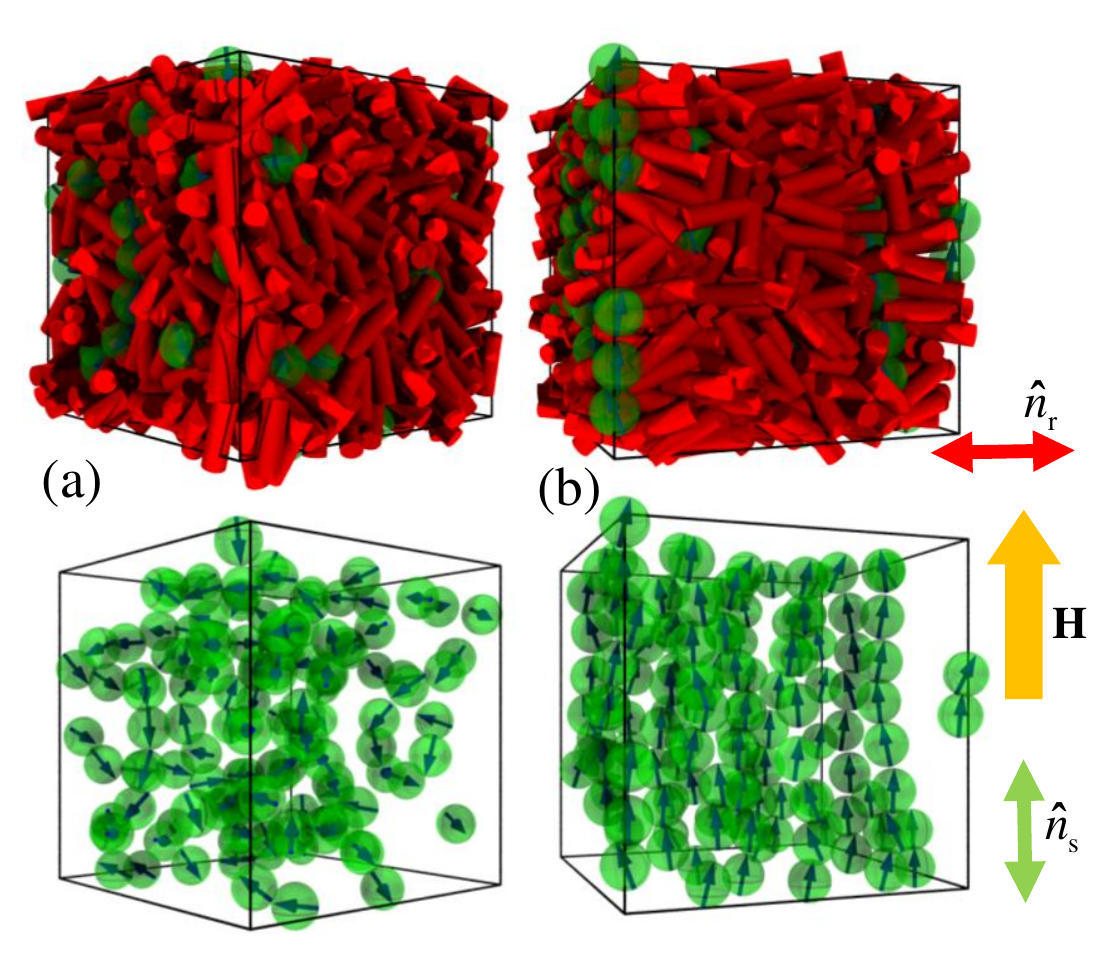}
  \caption{ Representative simulation snapshots of a GB-DSS mixture with $\sigma^{*}_{\text{s}}=2.0$ (a) in the isotropic state for $K=-0.9$ and $[(T^*,\rho^{*})=(2.4,0.338)]$ and (b) same state point subject to an external homogeneous magnetic field of strength $H^{*}=12.5$. The direction of the field and the directors $\hat{\mathbf{n}}_{\text{r}}$ and $\hat{\mathbf{n}}_{\text{s}}$ are sketched as arrows.}
  \label{fig:12}
  \end{figure}
\end{center}

We have also considered systems with larger magnetic spheres ($\sigma^{*}_{\text{s}}=2.0$) subject to an homogeneous magnetic field. More specifically, we have examined state points within the isotropic phase (at $T^{*}=2.6$ and $T^{*}=2.4$) and nematic phase (at $T^{*}=2.2$) at constant density $\mathrm{\rho}^{*}=0.338$ [see $(T^{*},\rho^{*})$ diagram in Fig \ref{fig:01}d]. The nematic and polar order parameters as a function of the field strength are given in Fig \ref{fig:11}; we use the parameter $K$ that is defined as $K=1-T^{*}/T^{*}_{\text{I}\text{N}_\text{u}}$ to indicate how far away the system temperature is from the temperature where the I-$\text{N}_\text{u}$ transformation occurs.  

Our results indicate (i) that nematic ordering is induced to the rodlike particles and (ii) the saturation values of the order parameters increase by decreasing the temperature of the system at constant density and (iii) the polar order parameters increase profoundly from zero (field off) and saturate to large values (over 0.8) even for small magnetic fields.  The field-dependence of the order parameter for the DSS, on the other hand, is less affected by the temperature reaching similar saturation values at strong fields.
 
One important finding is that the rodlike particles tend to be oriented perpendicular to the ferromagnetic chains (see Fig. \ref{fig:12}).  
Hence, 
a biaxial nematic order can be induced within an initially isotropic state subject to an homogeneous magnetic field.

\section{Conclusions}
In summary, using Monte-Carlo simulations we have studied the influence of dipolar interactions and external magnetic fields on the self-organization of binary mixtures of rods and soft spheres. 
We have implemented a tractable model of rods of Gay-Berne type (GB) and dipolar soft spheres (DSS) to explore these types of systems. Due to the simple nature of the modelled interactions and particle shapes such prototypes offer conclusive stereotypes for addressing fundamental issues regarding the structure of complex liquid crystalline ferrofluids. 

In the first part of the paper we investigated the underlying reference system, that is, a binary mixture of rods and purely repulsive soft spheres (SS). 
For this GB-SS mixture we obtain fully miscible isotropic, nematic, smectic (SmB) and lamellar phases. 
A central result of the present study is that the ratio $\sigma_{\text{s}}/\sigma_0$ (diameter of the spheres compared to the width of the rods) determines whether the system undergoes a disordered-ordered phase transition to the smectic ($\sigma_{\text{s}}/\sigma_0\approx 1$) or lamellar phase ($\sigma_s/\sigma_0\gtrsim 1.5$).  In the smectic phase the spherical particles organize into linear arrangements along the nematic director.  
Moreover, we find that the spheres are fitting in the  smectic layer  of rods where the side-by-side rod-sphere arrangement  
is preferred. By increasing the size of the spheres the smectic order is destroyed in favour of a lamellar phase.

The second part of the paper has been devoted to the impact of additional magnetic interactions on ordered states. To this end, we attached permanent magnetic moments to the centres of the soft spheres inducing strongly anisotropic long-ranged dipole-dipole interactions. 
For every configuration considered we observe a self-assembly of the dipolar spheres into chains  
due to a relatively large dipolar coupling ($\lambda \gtrsim 3.5$). The LC host stabilizes the orientational order of ferromagnetic chains. The ferromagnetic chains are randomly oriented up or down and the fluid has no macroscopic magnetization. 
Remarkably, the orientation of the dipolar spheres' director with respect to the nematic director of the rods depends sensitively on the interspecies size-to-width ratio. For $\sigma_{\text{s}}/\sigma_0\lesssim 1.5$ the ferromagnetic chains penetrate through the layers composed of (soft) rods forming an uniaxial smectic phase or at higher temperatures an uniaxial nematic phase (parallel directors) and for ratios $\sigma_{\text{s}}/\sigma_0\simeq 1.7$ the mixture forms a biaxial ferromagnetic lamellar phase (perpendicular directors).  We note that for $\sigma_{\text{s}}/\sigma_0\simeq 2.0$ we have also observed a biaxial nematic phase at intermediate temperatures (for details see Ref.\citenum{StavarXiv}).

As a further step, we have exposed the GB-DSS mixture to a (static) homogeneous magnetic field 
concentrating on isotropic states. Here we observe the reversed phenomenon, that is, the magnetic field induces orientational order to the chains (parallel to the field) that, in turn, imposes an orientational order to the rods. Depending on the size of the dipolar spheres the rods are oriented (on average) parallel ($\sigma_{\text{s}}/\sigma_0\simeq 1$) or perpendicular ($\sigma_{\text{s}}/\sigma_0\simeq 2$) to the field. 
The external field can thus be used to induce uniaxial or biaxial ferromagnetic order to a GB-DSS mixture which is isotropic in the absence of the field.

There are several directions which we believe require further investigation. 
In the present paper we considered only relatively small spheres ranging up to diameters double the width of the rods. 
What would happen in the presence of even smaller spheres? Besides an increased tendency of demixing into rod-rich and sphere-rich regions these systems could possibly exhibit new morphologies. Another interesting point concerns the self-assembly considered here with further types of external influences, e.g., time-dependent magnetic fields. 
In fact, a recent experiment on magnetic colloids under a rotating magnetic field \cite{Granick15} revealed a phase transition  from a disordered state to a layered state with hexagonal order; a phenomenon which was analyzed earlier in our group \cite{Sebastian11}. 
For the present system we expect that this field-induced structural transition imposes a nematic order where the rods are aligned between the magnetic layers. 
This could be a promising route for designing functional liquid crystalline ferrofluids with potential magneto-optical applications. 
We therefore hope that our results will stimulate further experimental investigations. 

\section{Acknowledgements}
We thank R. Stannarius and A. Eremin for stimulating discussions. Financial support from the German Science Foundation (DFG) via the priority programme
SPP 1681 is gratefully acknowledged.




\footnotesize{

\bibliographystyle{rsc} 

\begin{thebibliography}{49}
\providecommand*{\natexlab}[1]{#1}
\providecommand*{\mciteSetBstSublistMode}[1]{}
\providecommand*{\mciteSetBstMaxWidthForm}[2]{}
\providecommand*{\mciteBstWouldAddEndPuncttrue}
  {\def\EndOfBibitem{\unskip.}}
\providecommand*{\mciteBstWouldAddEndPunctfalse}
  {\let\EndOfBibitem\relax}
\providecommand*{\mciteSetBstMidEndSepPunct}[3]{}
\providecommand*{\mciteSetBstSublistLabelBeginEnd}[3]{}
\providecommand*{\EndOfBibitem}{}
\mciteSetBstSublistMode{f}
\mciteSetBstMaxWidthForm{subitem}
{(\emph{\alph{mcitesubitemcount}})}
\mciteSetBstSublistLabelBeginEnd{\mcitemaxwidthsubitemform\space}
{\relax}{\relax}

\bibitem[Asakura and Oosawa(1954)]{Asakura54}
S.~Asakura and F.~Oosawa, \emph{J. Chem. Phys.}, 1954, \textbf{22},
  1255--1256\relax
\mciteBstWouldAddEndPuncttrue
\mciteSetBstMidEndSepPunct{\mcitedefaultmidpunct}
{\mcitedefaultendpunct}{\mcitedefaultseppunct}\relax
\EndOfBibitem
\bibitem[Vliegenthart and Lekkerkerker(1999)]{Lekkerkerker99}
G.~A. Vliegenthart and H.~N.~W. Lekkerkerker, \emph{J. Chem. Phys.}, 1999,
  \textbf{111}, 4153--4157\relax
\mciteBstWouldAddEndPuncttrue
\mciteSetBstMidEndSepPunct{\mcitedefaultmidpunct}
{\mcitedefaultendpunct}{\mcitedefaultseppunct}\relax
\EndOfBibitem
\bibitem[Adams \emph{et~al.}(1998)Adams, Dogic, Keller, and Fraden]{Fraden98}
M.~Adams, Z.~Dogic, S.~L. Keller and S.~Fraden, \emph{Nature}, 1998,
  \textbf{393}, 349--352\relax
\mciteBstWouldAddEndPuncttrue
\mciteSetBstMidEndSepPunct{\mcitedefaultmidpunct}
{\mcitedefaultendpunct}{\mcitedefaultseppunct}\relax
\EndOfBibitem
\bibitem[Yasarawan and van Duijneveldt(2010)]{Jeroen10}
N.~Yasarawan and J.~S. van Duijneveldt, \emph{Soft Matter}, 2010, \textbf{6},
  353--362\relax
\mciteBstWouldAddEndPuncttrue
\mciteSetBstMidEndSepPunct{\mcitedefaultmidpunct}
{\mcitedefaultendpunct}{\mcitedefaultseppunct}\relax
\EndOfBibitem
\bibitem[Brader \emph{et~al.}(2002)Brader, Esztermann, and Schmidt]{Schmidt02}
J.~M. Brader, A.~Esztermann and M.~Schmidt, \emph{Phys. Rev. E}, 2002,
  \textbf{66}, 031401\relax
\mciteBstWouldAddEndPuncttrue
\mciteSetBstMidEndSepPunct{\mcitedefaultmidpunct}
{\mcitedefaultendpunct}{\mcitedefaultseppunct}\relax
\EndOfBibitem
\bibitem[{Peroukidis, S.D.} \emph{et~al.}(2010){Peroukidis, S.D.}, {Vanakaras,
  A.G.}, and {Photinos, D.J.}]{PeroukidisMix}
{Peroukidis, S.D.}, {Vanakaras, A.G.} and {Photinos, D.J.}, \emph{J. Mater.
  Chem.}, 2010, \textbf{20}, 10495--10502\relax
\mciteBstWouldAddEndPuncttrue
\mciteSetBstMidEndSepPunct{\mcitedefaultmidpunct}
{\mcitedefaultendpunct}{\mcitedefaultseppunct}\relax
\EndOfBibitem
\bibitem[Cinacchi \emph{et~al.}(2004)Cinacchi, Mederos, and
  Velasco]{Cinacchi04}
G.~Cinacchi, L.~Mederos and E.~Velasco, \emph{J. Chem. Phys.}, 2004,
  \textbf{121}, 3854–3867\relax
\mciteBstWouldAddEndPuncttrue
\mciteSetBstMidEndSepPunct{\mcitedefaultmidpunct}
{\mcitedefaultendpunct}{\mcitedefaultseppunct}\relax
\EndOfBibitem
\bibitem[Bolhuis \emph{et~al.}(2003)Bolhuis, Brader, and Schmidt]{Bolhuis03}
P.~G. Bolhuis, J.~M. Brader and M.~Schmidt, \emph{J. Phys. Condens. Matter},
  2003, \textbf{15}, S3421\relax
\mciteBstWouldAddEndPuncttrue
\mciteSetBstMidEndSepPunct{\mcitedefaultmidpunct}
{\mcitedefaultendpunct}{\mcitedefaultseppunct}\relax
\EndOfBibitem
\bibitem[Antypov and Cleaver(2004)]{Antypov04}
D.~Antypov and D.~J. Cleaver, \emph{J. Chem. Phys.}, 2004, \textbf{120},
  10307--10316\relax
\mciteBstWouldAddEndPuncttrue
\mciteSetBstMidEndSepPunct{\mcitedefaultmidpunct}
{\mcitedefaultendpunct}{\mcitedefaultseppunct}\relax
\EndOfBibitem
\bibitem[Urakami and Imai(2003)]{Urakami03}
N.~Urakami and M.~Imai, \emph{J. Chem. Phys.}, 2003, \textbf{119},
  2463–2470\relax
\mciteBstWouldAddEndPuncttrue
\mciteSetBstMidEndSepPunct{\mcitedefaultmidpunct}
{\mcitedefaultendpunct}{\mcitedefaultseppunct}\relax
\EndOfBibitem
\bibitem[Dogic \emph{et~al.}(2000)Dogic, Frenkel, and Fraden]{Dogic2000}
Z.~Dogic, D.~Frenkel and S.~Fraden, \emph{Phys. Rev. E}, 2000, \textbf{62},
  3925--3933\relax
\mciteBstWouldAddEndPuncttrue
\mciteSetBstMidEndSepPunct{\mcitedefaultmidpunct}
{\mcitedefaultendpunct}{\mcitedefaultseppunct}\relax
\EndOfBibitem
\bibitem[{Brochard, F.} and {de Gennes, P.G.}(1970)]{degennes70}
{Brochard, F.} and {de Gennes, P.G.}, \emph{J. Phys. France}, 1970,
  \textbf{31}, 691–708\relax
\mciteBstWouldAddEndPuncttrue
\mciteSetBstMidEndSepPunct{\mcitedefaultmidpunct}
{\mcitedefaultendpunct}{\mcitedefaultseppunct}\relax
\EndOfBibitem
\bibitem[Chen and Amer(1983)]{Chen83}
S.-H. Chen and N.~M. Amer, \emph{Phys. Rev. Lett.}, 1983, \textbf{51},
  2298--2301\relax
\mciteBstWouldAddEndPuncttrue
\mciteSetBstMidEndSepPunct{\mcitedefaultmidpunct}
{\mcitedefaultendpunct}{\mcitedefaultseppunct}\relax
\EndOfBibitem
\bibitem[Mirantsev(2014)]{Mirantsev14}
L.~V. Mirantsev, \emph{Phys. Lett. A}, 2014, \textbf{378}, 86–89\relax
\mciteBstWouldAddEndPuncttrue
\mciteSetBstMidEndSepPunct{\mcitedefaultmidpunct}
{\mcitedefaultendpunct}{\mcitedefaultseppunct}\relax
\EndOfBibitem
\bibitem[Mertelj \emph{et~al.}(2014)Mertelj, Osterman, Lisjak, and
  Copic]{Mertelj14}
A.~Mertelj, N.~Osterman, D.~Lisjak and M.~Copic, \emph{Soft Matter}, 2014,
  \textbf{10}, 9065–9072\relax
\mciteBstWouldAddEndPuncttrue
\mciteSetBstMidEndSepPunct{\mcitedefaultmidpunct}
{\mcitedefaultendpunct}{\mcitedefaultseppunct}\relax
\EndOfBibitem
\bibitem[Kreuzer \emph{et~al.}(1992)Kreuzer, Tschudi, and
  Eidenschink]{Kreuzer92}
M.~Kreuzer, T.~Tschudi and R.~Eidenschink, \emph{Mol. Cryst. Liq. Crys. A},
  1992, \textbf{223}, 219–227\relax
\mciteBstWouldAddEndPuncttrue
\mciteSetBstMidEndSepPunct{\mcitedefaultmidpunct}
{\mcitedefaultendpunct}{\mcitedefaultseppunct}\relax
\EndOfBibitem
\bibitem[Kreuzer \emph{et~al.}(1993)Kreuzer, Tschudi, de~Jeu, and
  Eidenschink]{Kreuzer93}
M.~Kreuzer, T.~Tschudi, W.~H. de~Jeu and R.~Eidenschink, \emph{Appl. Phys.
  Lett.}, 1993, \textbf{62}, 1712–1714\relax
\mciteBstWouldAddEndPuncttrue
\mciteSetBstMidEndSepPunct{\mcitedefaultmidpunct}
{\mcitedefaultendpunct}{\mcitedefaultseppunct}\relax
\EndOfBibitem
\bibitem[Kop\ifmmode~\check{c}\else \v{c}\fi{}ansk\'y
  \emph{et~al.}(2008)Kop\ifmmode~\check{c}\else \v{c}\fi{}ansk\'y, Toma\ifmmode
  \check{s}\else \v{s}\fi{}ovi\ifmmode~\check{c}\else \v{c}\fi{}ov\'a,
  Konerack\'a, Z\'avi\ifmmode~\check{s}\else \v{s}\fi{}ov\'a, Timko,
  D\ifmmode~\check{z}\else \v{z}\fi{}arov\'a, \ifmmode~\check{S}\else
  \v{S}\fi{}princov\'a, \'Eber, Fodor-Csorba, T\'oth-Katona, Vajda, and
  Jadzyn]{Kopacansky2008}
P.~Kop\ifmmode~\check{c}\else \v{c}\fi{}ansk\'y, N.~Toma\ifmmode \check{s}\else
  \v{s}\fi{}ovi\ifmmode~\check{c}\else \v{c}\fi{}ov\'a, M.~Konerack\'a,
  V.~Z\'avi\ifmmode~\check{s}\else \v{s}\fi{}ov\'a, M.~Timko, A.~c.~v.
  D\ifmmode~\check{z}\else \v{z}\fi{}arov\'a, A.~\ifmmode~\check{S}\else
  \v{S}\fi{}princov\'a, N.~\'Eber, K.~Fodor-Csorba, T.~T\'oth-Katona, A.~Vajda
  and J.~Jadzyn, \emph{Phys. Rev. E}, 2008, \textbf{78}, 011702\relax
\mciteBstWouldAddEndPuncttrue
\mciteSetBstMidEndSepPunct{\mcitedefaultmidpunct}
{\mcitedefaultendpunct}{\mcitedefaultseppunct}\relax
\EndOfBibitem
\bibitem[Buluy \emph{et~al.}(2011)Buluy, Nepijko, Reshetnyak, Ouskova,
  Zadorozhnii, Leonhardt, Ritschel, Schonhense, and Reznikov]{Buluy2011}
O.~Buluy, S.~Nepijko, V.~Reshetnyak, E.~Ouskova, V.~Zadorozhnii, A.~Leonhardt,
  M.~Ritschel, G.~Schonhense and Y.~Reznikov, \emph{Soft Matter}, 2011,
  \textbf{7}, 644--649\relax
\mciteBstWouldAddEndPuncttrue
\mciteSetBstMidEndSepPunct{\mcitedefaultmidpunct}
{\mcitedefaultendpunct}{\mcitedefaultseppunct}\relax
\EndOfBibitem
\bibitem[Kredentser \emph{et~al.}(2013)Kredentser, Buluy, Davidson, Dozov,
  Malynych, Reshetnyak, Slyusarenko, and Reznikov]{Kredentser13}
S.~Kredentser, O.~Buluy, P.~Davidson, I.~Dozov, S.~Malynych, V.~Reshetnyak,
  K.~Slyusarenko and Y.~Reznikov, \emph{Soft Matter}, 2013, \textbf{9},
  5061--5066\relax
\mciteBstWouldAddEndPuncttrue
\mciteSetBstMidEndSepPunct{\mcitedefaultmidpunct}
{\mcitedefaultendpunct}{\mcitedefaultseppunct}\relax
\EndOfBibitem
\bibitem[pri(Private communication with R. Stannarius group)]{privcomm}
Private communication with R. Stannarius group\relax
\mciteBstWouldAddEndPuncttrue
\mciteSetBstMidEndSepPunct{\mcitedefaultmidpunct}
{\mcitedefaultendpunct}{\mcitedefaultseppunct}\relax
\EndOfBibitem
\bibitem[May \emph{et~al.}(2014)May, Stannarius, Klein, and Eremin]{May14}
K.~May, R.~Stannarius, S.~Klein and A.~Eremin, \emph{Langmuir}, 2014,
  \textbf{30}, 7070--7076\relax
\mciteBstWouldAddEndPuncttrue
\mciteSetBstMidEndSepPunct{\mcitedefaultmidpunct}
{\mcitedefaultendpunct}{\mcitedefaultseppunct}\relax
\EndOfBibitem
\bibitem[{Kyrylyuk Andriy V.} \emph{et~al.}(2011){Kyrylyuk Andriy V.}, {Hermant
  Marie Claire}, {Schilling Tanja}, {Klumperman Bert}, {Koning Cor E.}, and
  van~der Schoot~Paul]{Schoot11}
{Kyrylyuk Andriy V.}, {Hermant Marie Claire}, {Schilling Tanja}, {Klumperman
  Bert}, {Koning Cor E.} and van~der Schoot~Paul, \emph{Nat. Nanotechnol.},
  2011, \textbf{6}, 364--369\relax
\mciteBstWouldAddEndPuncttrue
\mciteSetBstMidEndSepPunct{\mcitedefaultmidpunct}
{\mcitedefaultendpunct}{\mcitedefaultseppunct}\relax
\EndOfBibitem
\bibitem[Klapp(2005)]{Klapp05}
S.~H.~L. Klapp, \emph{J. Phys. Condens. Matter}, 2005, \textbf{17}, R525\relax
\mciteBstWouldAddEndPuncttrue
\mciteSetBstMidEndSepPunct{\mcitedefaultmidpunct}
{\mcitedefaultendpunct}{\mcitedefaultseppunct}\relax
\EndOfBibitem
\bibitem[Sreekumari and Ilg(2013)]{Ilg13}
A.~Sreekumari and P.~Ilg, \emph{Phys. Rev. E}, 2013, \textbf{88}, 042315\relax
\mciteBstWouldAddEndPuncttrue
\mciteSetBstMidEndSepPunct{\mcitedefaultmidpunct}
{\mcitedefaultendpunct}{\mcitedefaultseppunct}\relax
\EndOfBibitem
\bibitem[Rovigatti \emph{et~al.}(2011)Rovigatti, Russo, and
  Sciortino]{Sciortino11}
L.~Rovigatti, J.~Russo and F.~Sciortino, \emph{Phys. Rev. Lett.}, 2011,
  \textbf{107}, 237801\relax
\mciteBstWouldAddEndPuncttrue
\mciteSetBstMidEndSepPunct{\mcitedefaultmidpunct}
{\mcitedefaultendpunct}{\mcitedefaultseppunct}\relax
\EndOfBibitem
\bibitem[Rovigatti \emph{et~al.}(2012)Rovigatti, Russo, and
  Sciortino]{Sciortino12}
L.~Rovigatti, J.~Russo and F.~Sciortino, \emph{Soft Matter}, 2012, \textbf{8},
  6310--6319\relax
\mciteBstWouldAddEndPuncttrue
\mciteSetBstMidEndSepPunct{\mcitedefaultmidpunct}
{\mcitedefaultendpunct}{\mcitedefaultseppunct}\relax
\EndOfBibitem
\bibitem[Martin \emph{et~al.}(1998)Martin, Anderson, and Tigges]{Martin98}
J.~E. Martin, R.~A. Anderson and C.~P. Tigges, \emph{J. Chem. Phys.}, 1998,
  \textbf{108}, 7887--7900\relax
\mciteBstWouldAddEndPuncttrue
\mciteSetBstMidEndSepPunct{\mcitedefaultmidpunct}
{\mcitedefaultendpunct}{\mcitedefaultseppunct}\relax
\EndOfBibitem
\bibitem[J\"ager and Klapp(2011)]{Sebastian11}
S.~J\"ager and S.~H.~L. Klapp, \emph{Soft Matter}, 2011, \textbf{7},
  6606--6616\relax
\mciteBstWouldAddEndPuncttrue
\mciteSetBstMidEndSepPunct{\mcitedefaultmidpunct}
{\mcitedefaultendpunct}{\mcitedefaultseppunct}\relax
\EndOfBibitem
\bibitem[{Douglas Jack F.}(2010)]{Douglas10}
{Douglas Jack F.}, \emph{Nature}, 2010, \textbf{463}, 302--303\relax
\mciteBstWouldAddEndPuncttrue
\mciteSetBstMidEndSepPunct{\mcitedefaultmidpunct}
{\mcitedefaultendpunct}{\mcitedefaultseppunct}\relax
\EndOfBibitem
\bibitem[Odenbach(2002)]{odenbachbook}
S.~Odenbach, \emph{Magnetoviscous effects in ferrofluids}, Springer, 2002\relax
\mciteBstWouldAddEndPuncttrue
\mciteSetBstMidEndSepPunct{\mcitedefaultmidpunct}
{\mcitedefaultendpunct}{\mcitedefaultseppunct}\relax
\EndOfBibitem
\bibitem[Plouffe \emph{et~al.}(2015)Plouffe, Murthy, and Lewis]{Lewis15}
B.~D. Plouffe, S.~K. Murthy and L.~H. Lewis, \emph{Rep. Prog. Phys.}, 2015,
  \textbf{78}, 016601\relax
\mciteBstWouldAddEndPuncttrue
\mciteSetBstMidEndSepPunct{\mcitedefaultmidpunct}
{\mcitedefaultendpunct}{\mcitedefaultseppunct}\relax
\EndOfBibitem
\bibitem[Wu \emph{et~al.}(2012)Wu, Wu, Lin, He, and Li]{Wu12}
Y.~Wu, Z.~Wu, X.~Lin, Q.~He and J.~Li, \emph{ACS Nano}, 2012, \textbf{6},
  10910--10916\relax
\mciteBstWouldAddEndPuncttrue
\mciteSetBstMidEndSepPunct{\mcitedefaultmidpunct}
{\mcitedefaultendpunct}{\mcitedefaultseppunct}\relax
\EndOfBibitem
\bibitem[Peroukidis and Klapp(2015)]{StavarXiv}
S.~D. Peroukidis and S.~H.~L. Klapp, \emph{arXiv:1503.08277}, 2015\relax
\mciteBstWouldAddEndPuncttrue
\mciteSetBstMidEndSepPunct{\mcitedefaultmidpunct}
{\mcitedefaultendpunct}{\mcitedefaultseppunct}\relax
\EndOfBibitem
\bibitem[Gay and Berne(1981)]{GayBerne81}
J.~G. Gay and B.~J. Berne, \emph{J. Chem. Phys.}, 1981, \textbf{74},
  3316--3319\relax
\mciteBstWouldAddEndPuncttrue
\mciteSetBstMidEndSepPunct{\mcitedefaultmidpunct}
{\mcitedefaultendpunct}{\mcitedefaultseppunct}\relax
\EndOfBibitem
\bibitem[Berne and Pechukas(1972)]{BernePechukas72}
B.~J. Berne and P.~Pechukas, \emph{J. Chem. Phys.}, 1972, \textbf{56},
  4213--4216\relax
\mciteBstWouldAddEndPuncttrue
\mciteSetBstMidEndSepPunct{\mcitedefaultmidpunct}
{\mcitedefaultendpunct}{\mcitedefaultseppunct}\relax
\EndOfBibitem
\bibitem[Allen and Tildesley(2006)]{allen}
P.~Allen and D.~Tildesley, \emph{Computer Simulation of Liquids}, Oxford
  University Press, 2006\relax
\mciteBstWouldAddEndPuncttrue
\mciteSetBstMidEndSepPunct{\mcitedefaultmidpunct}
{\mcitedefaultendpunct}{\mcitedefaultseppunct}\relax
\EndOfBibitem
\bibitem[Cleaver \emph{et~al.}(1996)Cleaver, Care, Allen, and Neal]{Cleaver96}
D.~J. Cleaver, C.~M. Care, M.~P. Allen and M.~P. Neal, \emph{Phys. Rev. E},
  1996, \textbf{54}, 559--567\relax
\mciteBstWouldAddEndPuncttrue
\mciteSetBstMidEndSepPunct{\mcitedefaultmidpunct}
{\mcitedefaultendpunct}{\mcitedefaultseppunct}\relax
\EndOfBibitem
\bibitem[Schoen and Klapp(2007)]{Ewald07}
M.~Schoen and S.~Klapp, \emph{Reviews of Computational Chemistry}, 2007,
  \textbf{24}, 1--517\relax
\mciteBstWouldAddEndPuncttrue
\mciteSetBstMidEndSepPunct{\mcitedefaultmidpunct}
{\mcitedefaultendpunct}{\mcitedefaultseppunct}\relax
\EndOfBibitem
\bibitem[Camp \emph{et~al.}(1999)Camp, Allen, and Masters]{Camp1999}
P.~Camp, M.~Allen and A.~Masters, \emph{J. Chem. Phys.}, 1999, \textbf{101},
  9871\relax
\mciteBstWouldAddEndPuncttrue
\mciteSetBstMidEndSepPunct{\mcitedefaultmidpunct}
{\mcitedefaultendpunct}{\mcitedefaultseppunct}\relax
\EndOfBibitem
\bibitem[Cuetos \emph{et~al.}(2008)Cuetos, Galindo, and Jackson]{Cuetos2008}
A.~Cuetos, A.~Galindo and G.~Jackson, \emph{Phys. Rev. Lett.}, 2008,
  \textbf{101}, 237802\relax
\mciteBstWouldAddEndPuncttrue
\mciteSetBstMidEndSepPunct{\mcitedefaultmidpunct}
{\mcitedefaultendpunct}{\mcitedefaultseppunct}\relax
\EndOfBibitem
\bibitem[Veerman and Frenkel(1992)]{Veerman1992}
J.~Veerman and D.~Frenkel, \emph{Phys. Rev. A}, 1992, \textbf{45}, 5632\relax
\mciteBstWouldAddEndPuncttrue
\mciteSetBstMidEndSepPunct{\mcitedefaultmidpunct}
{\mcitedefaultendpunct}{\mcitedefaultseppunct}\relax
\EndOfBibitem
\bibitem[Berardi and Zannoni(2000)]{Bebo2000}
R.~Berardi and C.~Zannoni, \emph{J. Chem. Phys.}, 2000, \textbf{113},
  5971\relax
\mciteBstWouldAddEndPuncttrue
\mciteSetBstMidEndSepPunct{\mcitedefaultmidpunct}
{\mcitedefaultendpunct}{\mcitedefaultseppunct}\relax
\EndOfBibitem
\bibitem[McGrother \emph{et~al.}(1996)McGrother, Williamson, and
  Jackson]{Mcgrother1996}
S.~McGrother, D.~Williamson and G.~Jackson, \emph{J. Chem. Phys.}, 1996,
  \textbf{104}, 6755\relax
\mciteBstWouldAddEndPuncttrue
\mciteSetBstMidEndSepPunct{\mcitedefaultmidpunct}
{\mcitedefaultendpunct}{\mcitedefaultseppunct}\relax
\EndOfBibitem
\bibitem[De~Miguel \emph{et~al.}(1991)De~Miguel, Rull, Chalam, and
  Gubbins]{Miguel99}
E.~De~Miguel, L.~Rull, M.~Chalam and K.~Gubbins, \emph{Mol. Phys.}, 1991,
  \textbf{74}, 405--424\relax
\mciteBstWouldAddEndPuncttrue
\mciteSetBstMidEndSepPunct{\mcitedefaultmidpunct}
{\mcitedefaultendpunct}{\mcitedefaultseppunct}\relax
\EndOfBibitem
\bibitem[Klokkenburg \emph{et~al.}(2006)Klokkenburg, Erne, Meeldijk,
  Wiedenmann, Petukhov, Pullens, and Philipse]{Klokkenburg06}
M.~Klokkenburg, B.~Erne, J.~Meeldijk, A.~Wiedenmann, A.~Petukhov, R.~Pullens
  and A.~Philipse, \emph{Phys. Rev. Lett.}, 2006, \textbf{97}, 185702\relax
\mciteBstWouldAddEndPuncttrue
\mciteSetBstMidEndSepPunct{\mcitedefaultmidpunct}
{\mcitedefaultendpunct}{\mcitedefaultseppunct}\relax
\EndOfBibitem
\bibitem[Borb\'ath \emph{et~al.}(2014)Borb\'ath, Borb\'ath, G\"unther,
  Marinica, V\'ek\'as, , and Odenbach]{Odenbach2014}
T.~Borb\'ath, I.~Borb\'ath, S.~G\"unther, O.~Marinica, L.~V\'ek\'as,  and
  S.~Odenbach, \emph{Smart Mater. Struct.}, 2014, \textbf{23}, 055018\relax
\mciteBstWouldAddEndPuncttrue
\mciteSetBstMidEndSepPunct{\mcitedefaultmidpunct}
{\mcitedefaultendpunct}{\mcitedefaultseppunct}\relax
\EndOfBibitem
\bibitem[Jordanovic \emph{et~al.}(2011)Jordanovic, J\"ager, and
  Klapp]{Jordan2011}
J.~Jordanovic, S.~J\"ager and S.~Klapp, \emph{Phys. Rev. Lett.}, 2011,
  \textbf{106}, 038301\relax
\mciteBstWouldAddEndPuncttrue
\mciteSetBstMidEndSepPunct{\mcitedefaultmidpunct}
{\mcitedefaultendpunct}{\mcitedefaultseppunct}\relax
\EndOfBibitem
\bibitem[Yan \emph{et~al.}(2015)Yan, Bae, and Granick]{Granick15}
J.~Yan, S.~C. Bae and S.~Granick, \emph{Soft Matter}, 2015, \textbf{11},
  147--153\relax
\mciteBstWouldAddEndPuncttrue
\mciteSetBstMidEndSepPunct{\mcitedefaultmidpunct}
{\mcitedefaultendpunct}{\mcitedefaultseppunct}\relax
\EndOfBibitem
\end{thebibliography}
}

\end{document}